%% file: 00-root.tex
\documentclass[sigconf,natbib=true]{acmart}
\usepackage{booktabs} 
\usepackage{graphicx}
\usepackage{fdiaz}
\usepackage{algpseudocode}
\usepackage{algorithm}
\usepackage{dsfont}
\usepackage{amsmath}
\usepackage[english]{babel}
\newtheorem{theorem}{Theorem}
\newtheorem{corollary}{Corollary}
\usepackage{multirow} 
\usepackage{subcaption}
\usepackage{array}
\usepackage{bbding}
\usepackage{xspace}
\usepackage{balance}
\usepackage{colortbl} 

\AtBeginDocument{%
  }

\copyrightyear{2025}
\acmYear{2025}
\setcopyright{cc}
\setcctype{by}
\acmConference[ICTIR '25]{Proceedings of the 2025 International ACM SIGIR Conference on Innovative Concepts and Theories in Information Retrieval (ICTIR)}{July 18, 2025}{Padua, Italy}
\acmBooktitle{Proceedings of the 2025 International ACM SIGIR Conference on Innovative Concepts and Theories in Information Retrieval (ICTIR) (ICTIR '25), July 18, 2025, Padua, Italy}\acmDOI{10.1145/3731120.3744599}
\acmISBN{979-8-4007-1861-8/2025/07}


\begin{document}
\title{Towards Fair RAG: On the Impact of Fair Ranking in Retrieval-Augmented Generation}

\author{To Eun Kim}
\affiliation{%
  \institution{Carnegie Mellon University}
  \city{Pittsburgh}
  \state{PA}
  \country{USA}
}
\email{toeunk@cs.cmu.edu}

\author{Fernando Diaz}
\affiliation{%
  \institution{Carnegie Mellon University}
  \city{Pittsburgh}
  \state{PA}
  \country{USA}
}
\email{diazf@acm.org}

\input{98-fairag-notation}

\input{98-new-commands}

\input{01-abstract}

\maketitle
\input{01-introduction}

\input{03-relatedwork.tex}
\input{05-methods.tex}
\input{06-experiments}
\input{07-results}
\input{08-discussion}

\input{09-conclusion}

\begin{acks}
This work was supported by NSF grant 2402874. 
Any opinions, findings and conclusions or recommendations expressed in this material are those of the authors and do not necessarily reflect those of the sponsors.
\end{acks}

\bibliographystyle{ACM-Reference-Format}
\balance
\bibliography{99-fairness-ref, 99-rag-ref}

\newpage
\appendix
\input{97-appendix}

\end{document}

%% file: 98-fairag-notation.tex
%


\newcommand{\instance}{x}
\newcommand{\inputSpace}{\mathcal{X}}
\newcommand{\target}{y}
\newcommand{\outputSpace}{\mathcal{Y}}

\newcommand{\rmodel}{\mathcal{R}}
\newcommand{\srmodel}{\mathcal{S}}
\newcommand{\mlmodel}{\mathcal{G}}

\newcommand{\doc}{d}
\newcommand{\query}{q}

\newcommand{\retrievalScores}{\mathbf{s}}

\newcommand{\metric}{\mu}
\newcommand{\utilityMetric}{\mu_u}
\newcommand{\fairnessMetric}{\mu_f}
\newcommand{\relevanceMetric}{\mu_r}

\newcommand{\targetExposure}{\mathbf{\epsilon}^{*}}
\newcommand{\expectedExposure}{\mathbf{\epsilon}}
\newcommand{\stringUtility}{u}

\newcommand{\collection}{\textrm{C}}
\newcommand{\rankedList}{\pi}
\newcommand{\rankedLists}{\sigma}
\newcommand{\rankOfdInL}{\rankedList(\doc)}

\newcommand{\worthiness}[2]{wor({#1}|{#2})}

\newcommand{\queryGenerator}{\phi_q}
\newcommand{\promptGenerator}{\phi_p}

\newcommand{\xbar}{\overline{\instance}}

\newcommand{\normEED}{\overline{\text{EE-D}}}
\newcommand{\normEER}{\overline{\text{EE-R}}}
\newcommand{\normEAED}{\overline{\text{EAE-D}}}
\newcommand{\normEU}{\overline{\text{EU}}}

%% file: 98-new-commands.tex
\newcommand{\flantfivexxl}{Flan-T5-XXL\xspace}
\newcommand{\flantfive}{Flan-T5-Base\xspace}
\newcommand{\flantfivesmall}{Flan-T5-Small\xspace}
\newcommand{\flanultwo}{Flan-UL2\xspace}

\newcommand{\pl}{Plackett-Luce\xspace}

\newcommand{\lamp}{LaMP\xspace}

\newcommand{\deterministicRAG}{\texttt{DeterministicRAG}\xspace}
\newcommand{\stochasticRAG}{\texttt{StochasticRAG}\xspace}

\newcommand{\norm}[1]{\left\lVert#1\right\rVert}
\newcommand{\ltwonorm}[1]{{\left\lVert#1\right\rVert}_{2}}

\newcommand\boldblue[1]{\textcolor{blue}{\textbf{#1}}}
\newcommand\blue[1]{\textcolor{blue}{#1}}

%% file: 01-abstract.tex
\begin{abstract}
Despite the central role of retrieval in retrieval-augmented generation (RAG) systems, much of the existing research on RAG overlooks the well-established field of fair ranking and fails to account for the interests of all stakeholders involved. In this paper, we conduct the first systematic evaluation of RAG systems that integrate fairness-aware rankings, addressing both ranking fairness and attribution fairness, which ensures equitable exposure of the sources cited in the generated content. Our evaluation focuses on measuring item-side fairness, specifically the fair exposure of relevant items retrieved by RAG systems, and investigates how this fairness impacts both the effectiveness of the systems and the attribution of sources in the generated output that users ultimately see. By experimenting with twelve RAG models across seven distinct tasks, we show that incorporating fairness-aware retrieval often maintains or even enhances both ranking quality and generation quality, countering the common belief that fairness compromises system performance. Additionally, we demonstrate that fair retrieval practices lead to more balanced attribution in the final responses, ensuring that the generator fairly cites the sources it relies on. Our findings underscore the importance of item-side fairness in retrieval and generation, laying the foundation for responsible and equitable RAG systems and guiding future research in fair ranking and attribution.
\end{abstract}

%% file: 01-introduction.tex
\section{Introduction}\label{sec:introduction}
\begin{figure}[t]
    \centering
    \resizebox{\columnwidth}{!}{ 
        \includegraphics[trim=210 210 160 220, clip]{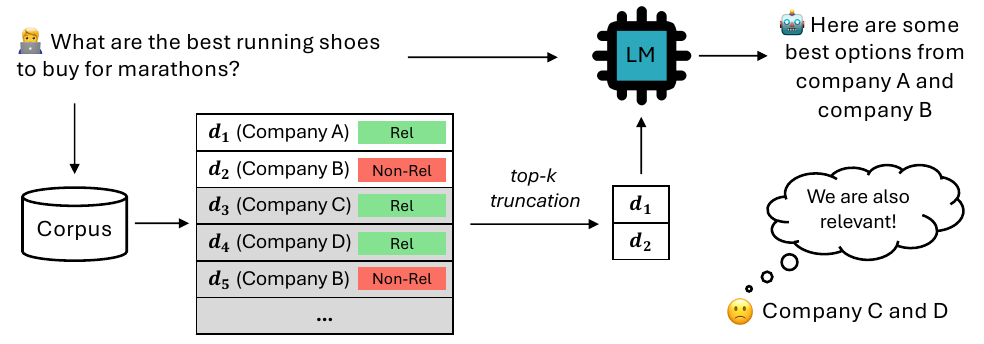}
    }
    \caption{Fairness concerns in RAG. A simplified example of how RAG models can ignore equally relevant items ($\doc_3$ and $\doc_4$) and always consume the fixed top-scoring items ($\doc_1$ and $\doc_2$) with the same order of ranking over the multiple user requests. This is due to the deterministic nature of the retrieval process and a short context-length of a language model that necessitates the top-$k$ truncation of a ranked list.}
    \label{fig:motivation}
    \vspace{-10pt}
\end{figure}

In recent years, the concept of fair ranking has emerged as a critical concern in modern information access systems \citep{ekstrand-diaz2022fairness-survey}. However, despite its significance, fair ranking has yet to be thoroughly examined in the context of retrieval-augmented generation (RAG) \citep{Lewis+al:2020, asai2024selfrag}, a rapidly advancing trend in natural language processing (NLP) systems \citep{kim2024reml}.
To understand why this is important, consider the RAG system in Figure \ref{fig:motivation}, where a user asks a question about running shoes. A classic retrieval system might return several documents containing information from various running shoe companies. If the RAG system only selects the top two documents, then information from the remaining two relevant companies will not be relayed to the predictive model and will likely be omitted from its answer.  The fair ranking literature refers to this situation as unfair because some relevant companies (i.e., in documents at position 3 and 4) receive less or no exposure compared to equally relevant company in the top position \citep{ekstrand-diaz2022fairness-survey}.

Understanding the effect of fair ranking in RAG is fundamental to ensuring responsible and equitable NLP systems.  Since retrieval results in RAG often underlie response attribution \citep{gao:alce}, \textit{unfair exposure of content} to the RAG system can result in incomplete evidence in responses (thus compromising recall of potentially relevant information for users) or downstream representational harms (thus creating or reinforcing biases across the set of relevant entities).  
In situations where content providers are compensated for contributions to inference, there can be financial implications for the unfairness \citep{balan:ekila,chen:resp-ai-gen,henderson:fm-fair-use}.  Indeed, the fair ranking literature indicates that these are precisely the harms that emerge when \textit{people} are searchers \citep{ekstrand-diaz2022fairness-survey}, much less RAG systems, where the searchers are \textit{machines}.  RAG complicates these challenges since it often truncates rankings to much shorter lengths to fit the generator's limited context size \citep{Bahri2020ChoppyCT, FiD-Light, kim2024reml}, making equal exposure of relevant items even harder.

Moreover, the fact that machines are now the searchers necessitates a different notion of item-worthiness (how deserving an item is to be included in a ranked list). Traditionally, ranking quality has been assessed based on relevance labels, which are created according to how relevant an item is to the user's query \citep{saracevic2016notion}. However, with RAG systems, where the consumer is a language model, there is a growing shift towards evaluating ranking quality based on \textit{utility labels}, which are determined by the usefulness of an item in aiding the model's task performance, rather than its relevance to the query \citep{salemi2024erag, Zhang24llmutilityjudgment}.

This shift from relevance to utility in the concept of item-worthiness can significantly alter our understanding of the relationship between fairness and ranking quality \citep{Balagopalan23roleofrelevance}---particularly the tradeoffs that are well-known in the fair ranking literature \citep{biega2018equity, diaz_2020_stochastic_ranking, singh_policylearning_fairness_2019}. Since previous fair ranking studies were conducted based on relevance judgments, they may need to be reexamined in light of utility-based judgments within the context of RAG.

However, purely focusing on how often certain items appear in the top-$k$ positions can neglect the fact that not all retrieved items are necessarily \textit{attributed} in the final generated response. If an item is retrieved but never actually influences the RAG model’s output, one cannot fully gauge whether it truly received exposure from the standpoint of the final generation. This reveals a subtle yet important gap: \textit{fair retrieval} may not directly translate to \textit{fair consumption}, especially when some retrieved items might be overshadowed in the generation step. Measuring how exposure is ultimately distributed across the attributed sources in the final response offers a more complete picture of exposure-based fairness in the context of RAG systems.

Our research aims to bridge the gap between traditional fair ranking studies and the emerging changes posed by RAG systems, ultimately enhancing our understanding of the interplay between fairness, ranking quality, and the effectiveness of RAG systems. We do this by evaluating RAG systems with a fairness-aware retriever across seven different tasks, experimenting with varying levels of retrieval fairness to observe changes in ranking quality and generation quality (utility)\footnote{Throughout this paper, we use "utility" and "generation quality" interchangeably to refer to the downstream effectiveness of RAG models, measured by arbitrary string utility metrics.}, as well as the fairness of attributed sources.

Our empirical results show that, in the context of machine users, there also exists an overall trend of fairness-quality tradeoff with respect to both retrieval and generation quality. However, the magnitude of this tradeoff is not particularly severe. In fact, we find that RAG models equipped with a fair ranker can often preserve a significant level of retrieval and generation quality, and in some cases, even surpass the quality achieved by the traditional RAG setup with a deterministic ranker that lacks fairness considerations.
Moreover, while the fraction of retrieved sources that actually appear in the final response may vary, equitable retrieval frequently leads to more equitable usage of those sources by the generator.

This surprising finding offers significant insight into the potential of RAG-based applications, suggesting that fair treatment of individual content providers can be achieved without sacrificing much of the high-quality service delivered to end-users. This challenges the conventional assumption of an inevitable tradeoff between fairness and quality, opening new avenues for developing more equitable and effective RAG systems.

%% file: 03-relatedwork.tex
\section{Background \& Related Work}
\subsection{Retrieval-Augmented Generation}
RAG, a specific type of retrieval-enhanced machine learning (REML) \citep{zamani:reml, kim2024reml}, has been widely adopted in various domains, including language modeling \citep{Khandelwal2020Generalization}, question-answering \citep{izacard_few-shot_2022}, and personalization \citep{salemi:lamp, salemi2024optimization, neelakanteswara2024ragstostyle}.
Studies on the evaluation of RAG models have primarily focused on their effectiveness, including end-to-end performance \citep{izacard_few-shot_2022, guu-realm, Lewis+al:2020} and the assessment of individual components \citep{es-etal-2024-ragas, saadfalcon2023ares, salemi2024erag, ru2024ragchecker}, such as retrieval relevance and model faithfulness.
Furthermore, recent efforts have explored attribution mechanisms for ensuring the trustworthiness of RAG responses \citep{gao:rarr, gao:alce}, examining how thoroughly models reference the source text and thus promoting a more faithful generation process.

However, little research has focused on evaluating fairness in retrieval-enhanced generation models, with the exception of recent work \citep{Shrestha_2024_CVPR}, which improved demographic diversity in human image generation by conditioning a generative model with externally retrieved images that help debias the generation process.

\subsection{Fairness in Ranking}
Fair ranking has been approached through various definitions based on normative concerns, primarily with distinctions made according to the stakeholders we prioritize. These include consumer-side fairness \citep{Mehrotra17auditing, ekstrand24notjustalgo}, which focuses on how fairly a system delivers satisfaction to users; provider-side fairness \citep{Sapiezynski19quantifying, JaenichSigir24FairReranking}, which addresses how fairly item providers receive monetary or reputational rewards; and item-side fairness \citep{Jiang24ItemsideLLM}, which examines how fairly items are treated in terms of representation or exposure. The motivation of item-side fairness is closely linked to provider-side fairness, as unfair treatment of items can lead to unfair compensation for providers. These fairness concerns can be further categorized by the scope of stakeholders, encompassing individual fairness---ensuring similar treatment for similar individuals---and group fairness---ensuring equitable outcomes across different groups \citep{caton2020fairness-survey, ekstrand-diaz2022fairness-survey}.
Previous studies have focused on developing metrics to measure fairness \citep{raj2020comparing-fair-metrics} and optimizing fair retrievers within a single \citep{yang2017measuring, zehlike2017fair, Sapiezynski19quantifying} or multiple rankings \citep{diaz_2020_stochastic_ranking, singh2018fairness, biega2018equity, singh_policylearning_fairness_2019}. In the context of provider- and item-side fairness, ensuring equal exposure of similar items across multiple rankings has gained significant attention \citep{ekstrand-diaz2022fairness-survey}. To achieve this, researchers have used stochastic rankers that return a distribution of rankings, in contrast to deterministic rankers commonly found in areas like RAG, which produce a fixed ranking. This approach ensures that, in expectation, similar items receive equal exposure across multiple user requests, with the distributions typically based on the merit of the rankings, such as an item's relevance \citep{diaz_2020_stochastic_ranking, singh_policylearning_fairness_2019}.

In this research, we employ a stochastic ranker in RAG to enhance \textit{individual item-side fairness}, aiming to ensure equal expected exposure for items that offer similar merits.

%% file: 05-methods.tex
\section{Experimental Methodology}\label{sec:methods}

In traditional RAG systems, a user input is used to query a retrieval system for recommended items from some corpus, which are then used for generation.  Given user input $\instance$, a query $\query$ generated by the query generation function $\queryGenerator(\instance)$, and a corpus of documents $\collection$, a \textit{deterministic retriever} $\rmodel(\query, \collection)$ returns a fixed ranked list $L$ every time $\query$ is seen. Retrieval is followed by a top $k$ truncation which is passed to a prompt generation function $\promptGenerator(\instance, L_{1:k})$ that returns a final prompt $\bar{\instance}$, which is subsequently passed to the language model $\mlmodel(\bar{\instance})$.
Because deterministic retrievers allocate exposure to the same item over repeated samples, RAG systems with deterministic retrievers present a challenge to ensuring equal exposure of relevant items to the generator.

\begin{figure}[]
    \centering
    \resizebox{\columnwidth}{!}{
    \includegraphics[trim=236 215 188 200, clip]{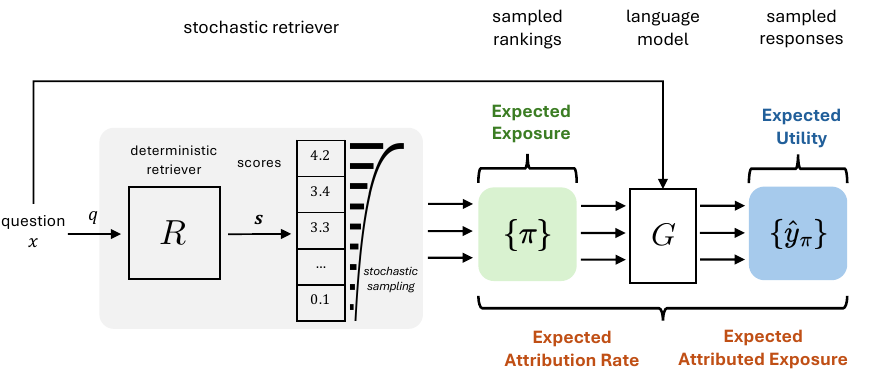}
    }
    \caption{Overview of our experimental design for examining how item-fairness in retrieval impacts both ranking and generation quality, as well as the fairness of attributed items in retrieval-augmented generation. To evaluate system performance across multiple identical user requests, we sample $N$ rankings from a stochastic retriever. Then, we measure the fairness and quality of these rankings (\emph{Expected Exposure} \S\ref{subsubsec:ee}), and then assess the system’s overall performance (\emph{Expected Utility} \S\ref{subsubsec:eu}), the usage of retrieved evidence (\emph{Expected Attribution Rate} \S\ref{subsubsec:ear}), and the fairness of attributed sources (\emph{Expected Attributed Exposure} \S\ref{subsubsec:eae}). The query and prompt generators are omitted for brevity.
    }
    \label{fig:exp-overview}
\end{figure}

To address the issue of unfairness in the rankings passed to the generator, we can convert a deterministic retriever into a stochastic retriever, which can, in expectation, provide fair rankings  \citep{diaz_2020_stochastic_ranking}. 
By sampling a ranking based on its quality to users---in this case generators---the expected exposure of different relevant items becomes similar and, therefore, fairer.  
Because decisions are stochastic, the fairness and quality of stochastic retrieval is evaluated based on a sample of rankings. Likewise, since each sampled ranking is processed by a generator, we measure the expected generator effectiveness over these sampled rankings. 
We also consider how fairly exposure is allocated among the items that end up being attributed by the final output, acknowledging that some retrieved items may not be credited in the generated text.
The complete evaluation pipeline of a RAG system with a stochastic retriever is illustrated in Figure \ref{fig:exp-overview}.

The following sections describe how we construct a test collection with utility labels (\S\ref{subsec:test-collection}), 
how we stochastically sample multiple rankings (\S\ref{subsec:methods:stochastic}), 
how we evaluate the fairness and ranking quality of the sampled rankings (\S\ref{subsubsec:ee}), 
and how we assess a RAG system’s performance over sampled rankings (\S\ref{subsubsec:eu}). 
We then explain how we measure the expected number of attributed items (\S\ref{subsubsec:ear}) 
and evaluate the fairness of these attributed sources (\S\ref{subsubsec:eae}).

\subsection{Construction of a Test Collection with Utility Labels}\label{subsec:test-collection}

Setting an appropriate proxy for measuring item-worthiness is crucial in the evaluation of fairness \citep{Balagopalan23roleofrelevance}. Drawing on the insight that utility-based judgments are more suitable than relevance judgments in the context of RAG \citep{Zhang24llmutilityjudgment, salemi2024erag}, we annotate item-level utility labels for all items in the corpus.

We define an item's worthiness by the marginal gain in utility (utility-gain) it provides to a language model (specifically, the generator in a RAG system) when used to solve a specific task as part of the augmentation process. To assess this utility-gain, each item in the corpus is individually supplied to the generator along with an input question. The utility-gain is then calculated as the difference between the utility of the augmented generator and that of a baseline language model without any information about the item.
Formally, let $\stringUtility_i$ denote the baseline string utility score from the vanilla language model prompted only with the input question, and let $\stringUtility_j$ represent the utility score from the language model with a prompt augmented by the $j$'th item $\doc_j$ in the corpus. The item $\doc_j$ is considered useful if the utility-gain $\delta_j = \stringUtility_j - \stringUtility_i$ is positive, and not useful otherwise.

Therefore, the item-level utility labels are designed to be both task- and generator-dependent, as the utility of each item varies depending on the task and the language model used. This labeling process also aligns with the principles of task-based information retrieval, where, in the context of \textit{human} searchers, document utility may vary on how the user expects to use the document \citep{nsf-workshop-human-ir}.

\subsection{Fairness-Aware Stochastic Retriever}\label{subsec:methods:stochastic}
Stochastic retrievers have been used for various purposes, such as optimization of retrieval models \citep{bruch_stochastic_LTR_2020, zamani24stochasticRAG, guiver_bayesian_PL_2009, oosterhuis2021PLRank}, as well as ensuring equitable exposure of items \citep{oosterhuis2022PLRank3, diaz_2020_stochastic_ranking, oosterhuis2021PLRank}. 
Many of these studies use \pl sampling \citep{plackett1975analysis} to achieve the stochasticity of retrieval. We follow the line of research and formally define how we derive a fairness-aware stochastic retriever through \pl sampling. To enhance sampling efficiency, we adopt the methodology of  \citet{oosterhuis2021PLRank}, and for controllable randomization, we utilize the approach proposed by \citet{diaz_2020_stochastic_ranking}.

Given $n$ items in a corpus $\collection$, a vector of retrieval scores $\retrievalScores \in \mathbb{R}^n$ can be obtained from $\rmodel(\query, \collection)$, which can be used to generate a ranked list $L$. We then min-max normalize retrieval scores to be in [0, 1] in order to construct a multinomial distribution over items \citep{biega2018equity}. The probability of an item $\doc$ being selected as the $i$'th item in a new ranking $\rankedList$ through \pl sampling is given by
\begin{equation}
p(\doc | L_{1:i-1}) = \frac{\text{exp}(\bar{\retrievalScores}_\doc) \mathds{1}[\doc \notin L_{1:i-1}]}{\sum_{\doc'\in\collection\backslash L_{1:i-1}} \text{exp}(\bar{\retrievalScores}_{\doc'})}
\end{equation}
\noindent
where $L_{1:i-1}$ is the partial ranking up to position $i-1$, $\bar{\retrievalScores}$ represents the normalized retrieval score vector, and $\bar{\retrievalScores}_\doc$ is the normalized score of item $\doc$.
Using this probability, we iteratively sample an item, set its probability to 0, renormalize the distribution, and repeat the process. The probability of generating a complete ranking is then given by the product of the placement probabilities for each item, i.e., $p(\rankedList | \query) = \prod^{n}_{i=1} p(\rankedList_i | \rankedList_{1:i-1})$.

This repeated sampling and renormalization process can be efficiently managed using the Gumbel-Softmax trick \citep{gumbel1954statistical, maddison2017gumbel}, which enables the sampling of rankings to be performed at the speed of sorting \citep{oosterhuis2021PLRank}. To do so, for each sampling iteration, we draw $U_i \sim \text{Uniform}(0, 1)$, followed by generating a Gumbel noise $G_i = -\log(-\log(U_i))$. The probability of each sampled ranking is then obtained by sorting the items based on their perturbed scores $\tilde{\retrievalScores}_{\doc_i} = \bar{\retrievalScores}_{\doc_i} + G_i$.

\subsubsection{\textbf{Controlling the Level of Fairness}}
Adjusting the level of randomization directly controls the degree of item-fairness, aligning with our goal to observe how varying levels of fairness in rankings affect the ranking and generation quality of a RAG model.
To obtain the controllability, we follow the work of \citet{diaz_2020_stochastic_ranking} and use a fairness control parameter $\alpha$. 
We apply the scalar $\alpha$ to each value in the normalized score vector $\bar{\retrievalScores}$ by raising each value to the power of $\alpha$.\footnote{We normalized the values to the range of [1, 2] instead of [0, 1]. The addition of 1 effectively serves the same purpose as adjusting a real-numbered $\alpha$. We chose this range to allow for an integer-valued $\alpha$.}
This process is done before the scores are passed to the sampling policy. 
Therefore, the modified sampling distribution is thus defined as:
\begin{equation}\label{eq:alpha-control}
p(\doc | L_{1:i-1}) = \frac{\text{exp}(\bar{\retrievalScores}^{\alpha}_\doc) \mathds{1}[\doc \notin L_{1:i-1}]}{\sum_{\doc'\in\collection\backslash L_{1:i-1}} \text{exp}(\bar{\retrievalScores}^{\alpha}_{\doc'})}
\end{equation}
\noindent
This implies that the sharpness of the sampling distribution is controlled by the $\alpha$. A higher $\alpha$ amplifies the probability of items with higher retrieval scores being sampled. Therefore, if multiple rankings are sampled by the stochastic retriever with high $\alpha$, it results in high disparity (i.e., item-side unfairness) of sampled rankings. At extreme, with considerably high $\alpha$, the procedure results in the identical rankings which is the behavior of a deterministic ranker (i.e., maximum item-unfairness). 
On the other hand, a lower $\alpha$ reduces the disparity of sampled rankings, making the exposure distribution fairer. At extreme, when $\alpha=0$, the sampling procedure becomes uniformly random and achieves the lowest disparity (i.e., maximum item-fairness) in the sampled rankings.

\subsection{Evaluation}\label{subsec:metrics}
As mentioned in Section \ref{sec:methods}, because we are dealing with stochastic retrievers, we need to measure the \textit{expected} behavior of the system.  
Let $\srmodel(\retrievalScores, N, k)$ be the stochastic sampler that samples a set of $N$ rankings $\rankedLists=\{\rankedList\}$, given the deterministic retrieval scores $\retrievalScores$, where each ranking $\rankedList$ is truncated to the size of $k$. 
From each ranking, we can get an output $\hat{\target}_\rankedList = \mlmodel(\promptGenerator(\instance, \rankedList))$.
With an arbitrary fairness metric $\fairnessMetric(\rankedLists)$ and a ranking quality metric $\relevanceMetric(\rankedLists)$ that takes a set of rankings as an input, we can measure the degree of fairness and ranking quality of the sampled rankings.
Similarly, an arbitrary string utility metric $\utilityMetric(y, \hat{\target}_\rankedList)$, such as ROUGE, can be used to assess an expected effectiveness of a RAG system by calculating the average of the $N$ metric scores. 

In this paper, based on the empirical investigation done by \citet{raj2020comparing-fair-metrics}, we use expected exposure disparity (EE-D) and expected exposure relevance (EE-R) \citep{diaz_2020_stochastic_ranking} as $\fairnessMetric$ and $\relevanceMetric$, respectively (\S\ref{subsubsec:ee}).
For $\utilityMetric$, we select the metric depending on the task, and we get the expectation of the utility of a RAG model which we call an expected utility (EU) (\S\ref{subsubsec:eu}).
Beyond these metrics, we also measure the expected rate of attributed items (EAR) in \S\ref{subsubsec:ear}, capturing how many retrieved items are ultimately used by the generator, and introduce the \emph{expected attributed exposure} (EAE) in \S\ref{subsubsec:eae}, which evaluates fairness specifically among the items that are actually attributed in the final output that can ultimately be displayed to human users.

\subsubsection{\textbf{Expected Exposure in the Context of Machine Users}}\label{subsubsec:ee}
Expected Exposure (EE) \citep{diaz_2020_stochastic_ranking} works by estimating the exposure of items across rankings (e.g., $\rankedLists$) created by a subject model, and comparing them with an optimal set of rankings that always satisfy the item-fairness.
To represent the attention over $n$ items in the corpus given by the consumer (generator in RAG), an $n\times1$ system exposure vector $\expectedExposure$ is created. This is then compared with an $n\times1$ target exposure vector $\targetExposure$, where it represents the exposure of items allocated by an oracle retriever that always rank useful items above non-useful ones \citep{diaz_2020_stochastic_ranking}.

With the system and target exposure vector $\expectedExposure\in\mathbb{R}^n$ and $\targetExposure\in\mathbb{R}^n$, we can get the difference between the two by the squared $l2$ distance:
\begin{equation}\label{eq:ee-loss}
\ltwonorm{\expectedExposure - \targetExposure}^{2} = \ltwonorm{\expectedExposure}^{2} -2\langle\expectedExposure, \targetExposure\rangle + \ltwonorm{\targetExposure}^{2}
\end{equation}
\noindent
This difference yields two metrics useful for fairness and ranking quality evaluation. $\ltwonorm{\expectedExposure}^{2}$ can be a measure for disparity of rankings (EE-D), and $\langle\expectedExposure, \targetExposure\rangle$ can be a measure of ranking quality (EE-R) by calculating the degree of alignment of system exposure to the target exposure (i.e., how much of the exposure is on useful items). Therefore, the higher the value of EE-D, the more unfair the set of rankings are, and the higher the value of EE-R, the closer the set of system rankings are to the optimal set of rankings with respect to the ranking quality.


The exposure of an item is calculated by modeling users' (e.g., generators in RAG) attention to each item in a ranking. For example, one can assume that the user is affected by position bias and gives attention following an exponential decay \citep{moffat2008rbp}.
However, these browsing models were developed for human-users not for machine-users, so we need a different user behavior model for generators in RAG. For the simplicity of the metric, and given recent efforts to reduce position bias and promote more even attention in machine-user settings \citep{he-etal-2024-never, hsieh-etal-2024-found}, we assume that the machine user consumes all items passed to the context with equal attention but pays zero attention to items placed after the $k$'th position due to top-$k$ truncation. This makes the user browsing model a step function parameterized by $k$.
In this work, a relevance-independent machine-user model (MU)  is set to the step function that reflects the behavior of \textit{top-k} truncation of RAG:
\begin{equation}\label{eq:um}
\text{MU}(i) =
\begin{cases}
  1 & \text{if } i \leq k \\
  0 & \text{otherwise}
\end{cases}
= \mathds{1}[i \leq k]
\end{equation}
\noindent
Given this machine user browsing model and a mapping from item index to its rank denoted as $\bar{\rankedList}_\doc$, a system exposure for each item $\doc$ is calculated as
\begin{equation}
\expectedExposure_\doc = \sum_{\rankedList \in S_n} p(\rankedList | \query) \text{MU}(\bar{\rankedList}_\doc)
\end{equation}
\noindent
and target exposures for a useful item $\doc$ and a unuseful item $\doc^{-}$ are calculated as
\begin{equation}
\begin{aligned}
\targetExposure_\doc &= \frac{1}{m} \sum^{m}_{i=1} \text{MU}(i) = 
\begin{cases}
    1 & \text{if } m \leq k \\
    \frac{k}{m} & \text{otherwise}
\end{cases}
\quad
\targetExposure_{\doc^{-}} =
\begin{cases}
    \frac{k-m}{n-m} & \text{if } m \leq k \\
    0 & \text{otherwise}
\end{cases}
\end{aligned}
\label{eq:target-exp}
\end{equation}
where $m$ is the number of useful items in the corpus of size $n$.

\subsubsection{\textbf{Expected Utility}}\label{subsubsec:eu}
Given the set of $N$ sampled rankings $\rankedLists$, we individually augment the generator with each ranking $\rankedList\in\rankedLists$, resulting in $N$ outputs from the generator. The utility of these outputs is then measured using an arbitrary string utility metric $\utilityMetric$. To determine the anticipated utility of a RAG model with fair rankings—represented by the tuple of a stochastic ranking sampler $\srmodel$ and a generator $\mlmodel$—we calculate the expected utility (EU) of the RAG system given an instance $\instance$.
\begin{align}
\text{EU}(\langle\srmodel,\mlmodel\rangle | \instance)
&= \mathbb{E}_{\rankedList \sim \srmodel}[\utilityMetric(\target, \hat{\target}_\rankedList)] 
 \\
&= \sum_{\rankedList\in S_n} p(\rankedList | \query) \utilityMetric(\target, \hat{\target}_\rankedList)
\approx \frac{1}{N} \sum_{\rankedList\in\rankedLists} \utilityMetric(\target, \hat{\target}_{\rankedList})\notag
\end{align}
\noindent
where $\hat{\target}_\rankedList$ is the prediction of a system given the ranking $\rankedList$, $S_n$ is the symmetric group of a ranked list $L$ from the deterministic retriever $\rmodel$, and $\sum_{\rankedList\in S_n} p(\rankedList | \query) = 1$.

\subsubsection{\textbf{Expected Attribution Rate}}\label{subsubsec:ear}
We use a natural language inference (NLI) model for attribution, reflecting its effectiveness in measuring faithfulness. \citet{honovich-2022-true} find that NLI outperforms other metrics in a meta-evaluation, leading works such as automated citation evaluation using NLI \cite{gao:alce}, which strongly correlates with human judgments. Building on the human evaluation framework by \citet{rashkin-2023-ais}, \citet{gao:rarr} introduced a new NLI-based metric to approximate human judgments, followed by \citet{bohnet2022attributed}. These studies collectively show that NLI-based methods capture attribution quality in a manner closely aligned with human assessments, making them a reliable choice for our evaluation.

Given a ranking $\rankedList \in \rankedLists$ of $k$ items and a predicted output $\hat{\target}_{\rankedList}$, 
we therefore measure item \emph{attribution rate (AR)} using
\begin{align}
\mu_{\text{AR}}\bigl(\rankedList, \hat{\target}_{\rankedList}\bigr)
\;=\;
\frac{1}{k} \sum_{\doc \in \rankedList}
\text{NLI}(\doc,\hat{\target}_{\rankedList})
\end{align}
\noindent
where $\text{NLI}(\doc,\hat{\target}_{\rankedList})\!=\!1$ if item \(\doc\) entails the output \(\hat{\target}_{\rankedList}\), and \(0\) otherwise, by a natural language inference model.
\footnote{
This metric is similar to the context utilization metric from \citet{ru2024ragchecker}, which also measures how effectively the generator leverages retrieved context. However, \citet{ru2024ragchecker} decompose the generated response into multiple claims and focus only on the proportion of relevant items, whereas our approach applies a single NLI check for each retrieved item.
}

Analogous to expected utility, we define expected attribution rate (EAR) of a RAG system as
\begin{align}
\text{EAR}(\langle \srmodel,\mlmodel\rangle |
\instance)
&= \mathbb{E}_{\rankedList \sim \srmodel}[
  \mu_{\text{AR}}(\rankedList, \hat{\target}_{\rankedList})
] \\
&= \sum_{\rankedList \in S_n} p(\rankedList | \query)
\mu_{\text{AR}}(\rankedList, \hat{\target}_{\rankedList})
\approx
\frac{1}{N}
\sum_{\rankedList \in \rankedLists}
\mu_{\text{AR}}(\rankedList, \hat{\target}_{\rankedList}).\notag
\end{align}

\subsubsection{\textbf{Expected Attributed Exposure}}\label{subsubsec:eae}
While EE measures retrieval fairness and relevance, it does not capture whether the retrieved items actually appear in the final generated output or how fairly exposure is allocated among those items once they surface after generation. To address this gap, we define the \emph{attributed exposure} vector $\expectedExposure^a \in \mathbb{R}^n$, where an 
attributed exposure for each item $\doc$ is calculated as
\begin{align}
\expectedExposure^a_\doc =
\sum_{\rankedList \in S_n}
p(\rankedList | \query)
\text{NLI}(\doc,\hat{\target}_{\rankedList}).
\end{align}
Similar to EE, a disparity measure \emph{expected attributed exposure} disparity (EAE-D) can be derived by measuring the squared $l_2$ distance of an attributed exposure vector, thus
\begin{align}
\label{eq:eae}
\text{EAE-D}(\langle \srmodel,\mlmodel\rangle |
\instance)
= \ltwonorm{\expectedExposure^a}^{2}.
\end{align}
Hence, EAE-D measures how fairly exposure is allocated across the items that are explicitly used, addressing the shortcoming of EE by directly linking exposure to items that influence the final output of a RAG system.

\subsubsection{\textbf{Normalization of Metrics}}
Since EAR is an average over the percentage of entailed items for each sampled ranking, its values lie on a consistent scale with [0, 1] range. However, the bounds of EE-D, EE-R, and EAE-D depend on the number of useful items in the corpus. 
Consequently, we apply normalization on a per-query basis by min-max scaling each metric according to its theoretical lower and upper bounds. We denote the normalized EE-D, EE-R, and EAE-D as
$\overline{\text{EE-D}}$, $\overline{\text{EE-R}}$, and $\overline{\text{EAE-D}}$ respectively.

However, theoretically determining the bounds of the expected utility (EU) of a RAG model is challenging. To address this, we normalized the EU by the model's empirical upper bound, the maximum observed utility across all runs of the experiment with the same generators. To approach the true upper bound, these runs include RAG models with an oracle retriever that consistently ranks useful items (i.e., those with positive utility labels) above non-useful ones, stochastically returning one of the $m!(n-m)!$ different rankings, where $m$ represents the number of useful items in the corpus. We denote the normalized EU as 
$\overline{\text{EU}}$,
which can be interpreted as the distance to the optimal utility.
%

%% file: 06-experiments.tex
\section{Experiments}\label{sec:experiments}

We choose the LaMP benchmark \citep{salemi:lamp} for our dataset. It assesses the personalization capability of language models through retrieval-augmentation of users' interaction history in a platform. LaMP includes various prediction tasks, such as classification, regression, and generation, and is well-suited for tasks where multiple items can be relevant/useful, unlike QA tasks with typically one or two provenance items. The retrieval items in LaMP have clear providers and consumers, aligning with our goal to ensure fairness for individual item providers. For example, in LaMP-1, retrieval items are academic papers, where exposure can increase citation counts for authors. In LaMP-4, retrieval items are news articles, where exposure can lead to monetary compensation for journalists.\footnote{Throughout this paper, we use LaMP-1 to represent classification and LaMP-4 to represent text generation in visualizations. All LaMP tasks were used in our experiments, and similar trends were observed.}
Due to the absence of a test set, we constructed a test collection as described in \S\ref{subsec:test-collection}, using the first thousand entries of a user-based development set. 
Then, we discarded entries that have only one useful item in the corpus, as it is unnecessary to concern item-fairness in that case. We release the test collection along with dataset statistics. 

\input{06-experiments/graphics/alpha-control}

We use BM25 (lexical retriever) \citep{Robertson1995OkapiBM25}, SPLADE (learned sparse retriever) \citep{splade}, and Contriever (bi-encoder dense retriever) \citep{izacard2022contriever} as deterministic retrievers providing retrieval scores to base the sampling on.  These models  represent commonly used retrievers in the RAG literature \citep{kim2024reml}.
We use a sampling size of $N = 100$ and a truncation size of $k = 5$.

For generation models, we use \flantfivesmall, \flantfive, \flantfivexxl \citep{chung2022scaling-flant5}, and \flanultwo \citep{tay2023ul}.
For decoding strategy, beam size is set to 4, and no sampling strategy is used. This is to ensure that stochasticity is only introduced to the retriever for controlled experiments.
With the three base retrievers and four generators, we configure twelve different RAG models and evaluate them on the seven LaMP tasks. Utility measurement of the generated strings follows the metrics used in the LaMP paper.
For NLI computations, a RoBERTa large model \cite{liu2019roberta} fine-tuned on the natural language inference task is used.\footnote{https://huggingface.co/FacebookAI/roberta-large-mnli}

We repeat the experiments with four different fairness control parameters $\alpha=1, 2, 4, 8$, which allows us to assess the utility of the RAG models with different levels of item-fairness.
From Figure \ref{fig:alpha-control}, we observe how effectively $\alpha$, described in the Equation \ref{eq:alpha-control}, controls the disparity of rankings. 
For example, when $\alpha$ is set to 4, we usually obtain a set of sampled rankings with $\overline{\text{EE-D}}$ mostly in the range of [0.5, 0.8], and when $\alpha$ is set to 8, we often get a set of sampled rankings with $\overline{\text{EE-D}}=1$.
%

We conducted paired (per-query) \(t\)-tests for $\overline{\text{EE-D}}$ across the different $\alpha$ values and found statistically significant differences \((p<0.01)\) in all 84 experimental conditions (\(12\) models \(\times\) \(7\) tasks). 


%% file: 06-experiments/graphics/alpha-control.tex
\begin{figure}[]
    \centering
    \begin{subfigure}[b]{0.49\columnwidth}
        \centering
        \includegraphics[trim=13 10 40 40, clip, width=\columnwidth]{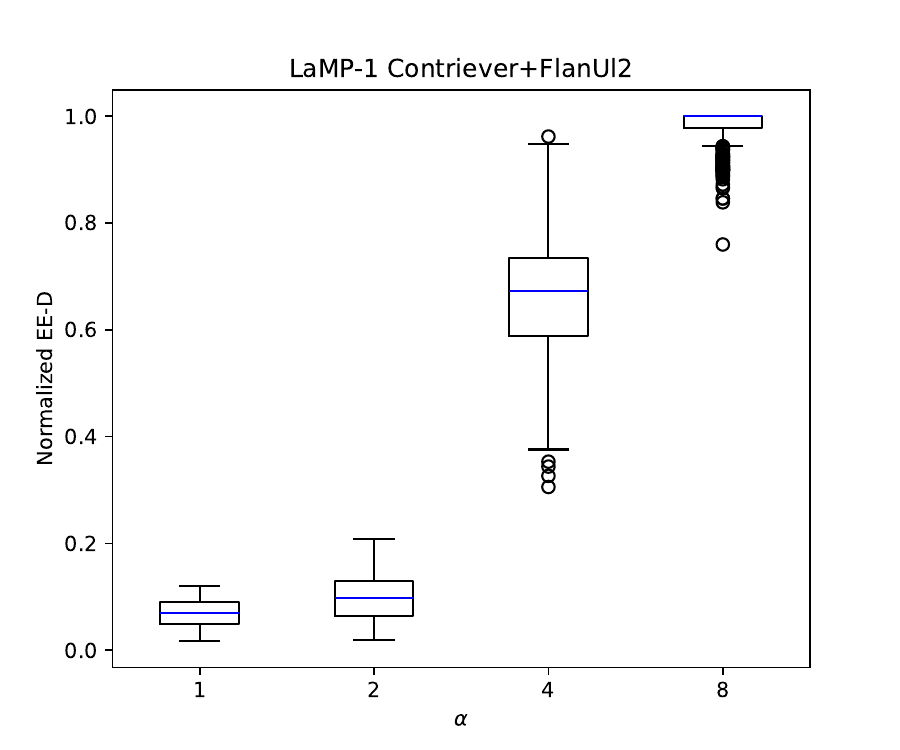}
        \caption{LaMP 1}
    \end{subfigure}
    \begin{subfigure}[b]{0.49\columnwidth}
        \centering
        \includegraphics[trim=13 10 40 40, clip, width=\columnwidth]{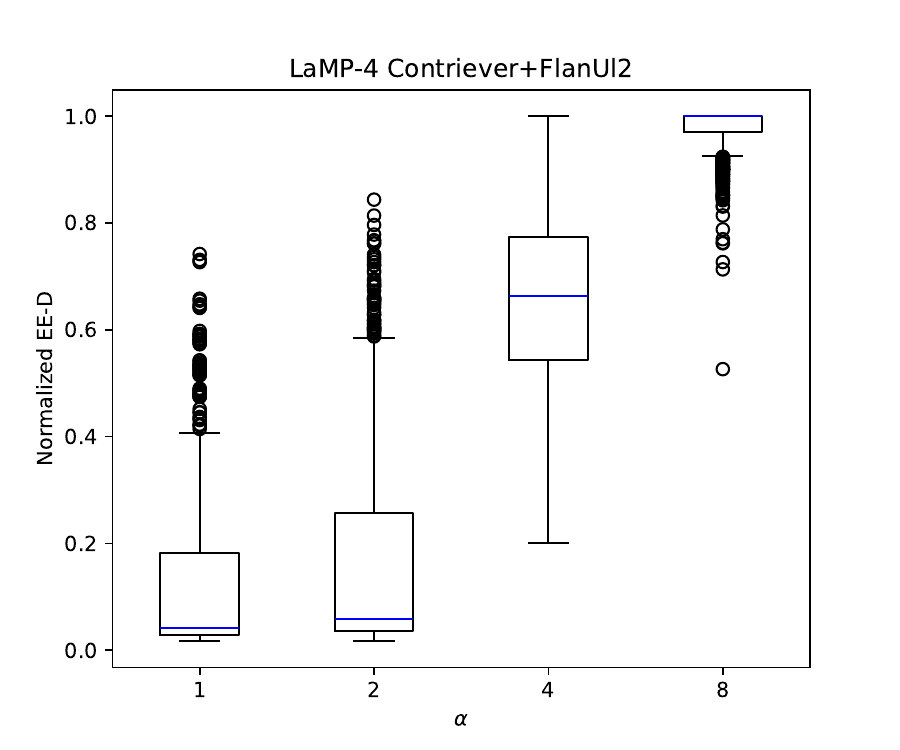}
        \caption{LaMP 4}
    \end{subfigure}
    \hfill
    \caption{Effect of a fairness control parameter ($\alpha$) on the disparity of rankings ($\overline{\text{EE-D}}$) in LaMP task 1 and 4. The RAG model is configured with the Contriever and \flanultwo. Each data point represents the normalized EE-D of each run of the experiment (i.e., one query $\rightarrow$ $N$ sampled rankings $\rightarrow$ $\overline{\text{EE-D}}$ of the $N$ rankings).
    }
    \label{fig:alpha-control}
    \vspace{-10pt}
\end{figure}

%% file: 07-results.tex
\section{Results}\label{sec:results}

\textit{\textbf{RQ1}: Is there a tradeoff between ensuring item-fairness in rankings and maintaining high ranking quality when utility labels are used for evaluation?}

By gathering all four repeated runs of the experiments with different $\alpha$ values, we can plot the trend of ranking quality ($\normEER$) against item fairness ($\normEED$), as shown in Figure \ref{subfig:eed-eer-curve}.

As shown in previous studies \citep{singh_policylearning_fairness_2019, diaz_2020_stochastic_ranking}, there is a well-known tradeoff between fairness and ranking quality for human users. Similarly, we observe a general tradeoff for machine users. However, unlike past findings, this tradeoff is not always strict. For instance, in Figure \ref{subfig:eed-eer-curve}, both SPLADE and Contriever maintain consistently high ranking quality while being considerably fairer, and for BM25, ranking quality even improves as fairness increases, up to a certain point.

At the rightmost side of the lines, where $\normEED = 1$ (representing the performance of deterministic rankers), we observe that these rankers do not always deliver the highest ranking quality. This suggests that commonly used deterministic rankers in RAG systems may be suboptimal, and that ranking quality can be improved while ensuring item fairness. This becomes even clearer when examining the impact of fair ranking on the downstream performance of a RAG system.

The leftmost side of the lines, where $\normEED = 0$, represents the performance of a uniformly random ranking policy. At this point, the measured ranking quality should approximate the proportion of positively labeled items in the corpus, which is 31\% for LaMP task 4.
This is notably higher than in non-RAG (human-user) settings, where the percentage of relevant documents is typically much smaller, resulting in a $\normEER$ value near 0 \citep{diaz_2020_stochastic_ranking}.

\input{07-discussion/tables/00-slope-auc}
To quantify the tradeoff, we fit a linear line to the experiment results, where a steeper slope reflects a stronger tradeoff between fairness and ranking quality (slope). We also quantify the performance of fair rankers, by calculating the area under the disparity-ranking quality curve (AUC; Figure \ref{subfig:eed-eer-curve}).
As can be seen in Table \ref{tab:slope-auc}, we observe that retriever that yields higher ranking quality has higher tradeoff in terms of retrieval fairness.

\input{07-discussion/figures/02-curve-lamp4-avg}
\input{07-discussion/tables/03-fairness-interval}

\textit{\textbf{RQ2}: Is there a tradeoff between ensuring item-fairness in ranking and maintaining high generation quality of a RAG model?}

Before examining the relationship between fairness and RAG utility, Figure \ref{subfig:eer-eu-curve} shows an auxiliary result confirming a strong correlation between utility-based ranking quality and the effectiveness of RAG models. This is unsurprising, as item-worthiness judgments were based on the utility-gain provided by the generator. However, this correlation suggests that the tradeoff observed in the disparity-ranking quality curve (Figure \ref{subfig:eed-eer-curve}) is likely to manifest similarly due to this strong relationship.

In fact, as observed from the disparity-utility curve (Figure \ref{subfig:eed-eu-curve}), we see a global trend of a non-strict tradeoff (i.e., RAG models maintain high generation quality while being considerably fair, and often even achieve higher quality).

However, a closer look at the local trend offers a significant insight: \textit{RAG systems with fair ranking can often achieve higher system-effectiveness compared to models with deterministic rankers}.
In Table \ref{tab:compare-baseline}, we divided the fairness levels into five disparity intervals based on the normalized EE-D. As shown in the table 
, improving fairness to the level of $\normEED \in [0.8, 1.0)$, and even $\normEED \in [0.6, 0.8)$, can often enhance the expected utility of many RAG models across LaMP tasks. 
For example, having $\normEED$ in the range of [0.8, 1.0) outperforms the baseline for all models in LaMP-4 and for eight out of twelve models in LaMP-1.

Significance testing results in Table \ref{tab:compare-baseline} provide additional nuance. The range
$\normEED \in [0.0, 0.2)$ shows a statistically significant drop in utility compared to the baseline, indicating that pushing disparity to extremely low levels can indeed reduce performance. Beyond this small-disparity range ($\normEED \in [0.2, 1.0)$), 
the differences in utility scores from the baseline either remain statistically insignificant or—when significant—reflect improved utility. This suggests that models can maintain a level of retrieval fairness close to the baseline’s utility performance without incurring a notable cost, and may even achieve higher effectiveness in many cases.

\input{07-discussion/figures/04-eae-lamp4}

\textit{\textbf{RQ3}: What is the impact of item-fairness in ranking to the fairness of the attributed sources used in the final response?}

To address this, we analyze three primary relationships: (1) how retrieval fairness ($\normEED$) relates to the expected attribution rate (EAR), (2) how ranking quality ($\normEER$) interacts with EAR, and (3) how retrieval fairness translates into consumption fairness ($\normEAED$) in the final output.

Figures \ref{subfig:eed-ear-curve} and \ref{subfig:eer-ear-curve} together shed light on how many items are actually attributed by the generator, which is a vital step in understanding how exposure might become more or less equitable once the generator has consumed the retrieved items. Figure \ref{subfig:eed-ear-curve} shows that when retrieval becomes more fair (lower $\normEED$), the generator tends to attribute fewer items overall, suggesting a tradeoff between distributing items equitably and actually getting them used in the final output. 

Interestingly, the general shape of $\normEED$ versus EAR in Figure \ref{subfig:eed-ear-curve} aligns closely with the shape of $\normEED$ versus $\normEER$ in Figure \ref{subfig:eed-eer-curve}. By looking at Figure \ref{subfig:eer-ear-curve}, it becomes clear that the generator is more likely to use multiple items if they are useful in solving the task, reflecting the idea that high-quality rankings encourage higher EAR. 


Figure \ref{subfig:eed-eaed-curve} directly addresses the question of whether fair retrieval leads to fair consumption by comparing $\normEED$ (\textit{retrieval fairness}) to $\normEAED$ (\textit{consumption fairness}). 
\footnote{Normalized EE-D and EAE-D scores are directly comparable, as they share the same dimensions in their exposure vectors and have the same $l1$ norm.}
The positive trend suggests that a more equitable ranking generally produces a more equitable distribution of attributed items in the final generation. However, the diagonal line (y=x) reveals subtle differences between fairness at the ranking stage and fairness at the consumption stage. Data points falling below this line imply that consumption is even fairer than the initial ranking might predict, while points above it suggest that the generator’s selective usage of items has introduced new disparity. This pattern shifts depending on how relevant and useful the retrieved items are, mirroring the earlier observation (Figure \ref{subfig:eer-ear-curve}) that the generator favors content that supports solving the task. In cases where many items have only marginal utility, the generator’s emphasis on a few high-utility sources can inadvertently raise disparity, placing the plotted values above the y=x line.

%% file: 07-discussion/tables/00-slope-auc.tex
\begin{table}[]
\centering
\begin{subtable}[t]{0.45\columnwidth}
    \centering
    \begin{tabular}{l|ll}
               & slope$\downarrow$  & AUC$\uparrow$    \\
               \hline
    BM25       & 0.1351 & 0.3603 \\
    SPLADE     & \textbf{0.1971} & \textbf{0.4166} \\
    Contriever & 0.1741 & 0.3864
    \end{tabular}
    \caption{LaMP-1}
\end{subtable}
\hfill
\begin{subtable}[t]{0.45\columnwidth}
    \centering
    \begin{tabular}{l|ll}
               & slope$\downarrow$  & AUC$\uparrow$    \\
               \hline
    BM25       & 0.0895 & 0.3921 \\
    SPLADE     & 0.1453 & 0.4205 \\
    Contriever & \textbf{0.1624} & \textbf{0.4287}
    \end{tabular}
    \caption{LaMP-4}
\end{subtable}
\caption{Comparison of fairness-ranking quality tradeoff (slope) and ranking quality (AUC) metrics for LaMP-1 and LaMP-4 datasets.}
\label{tab:slope-auc}
\vspace{-30pt}
\end{table}

%% file: 07-discussion/figures/02-curve-lamp4-avg.tex
\begin{figure*}[ht]
    \centering
    \begin{subfigure}[b]{0.3\textwidth}
        \centering
        \includegraphics[width=\textwidth]{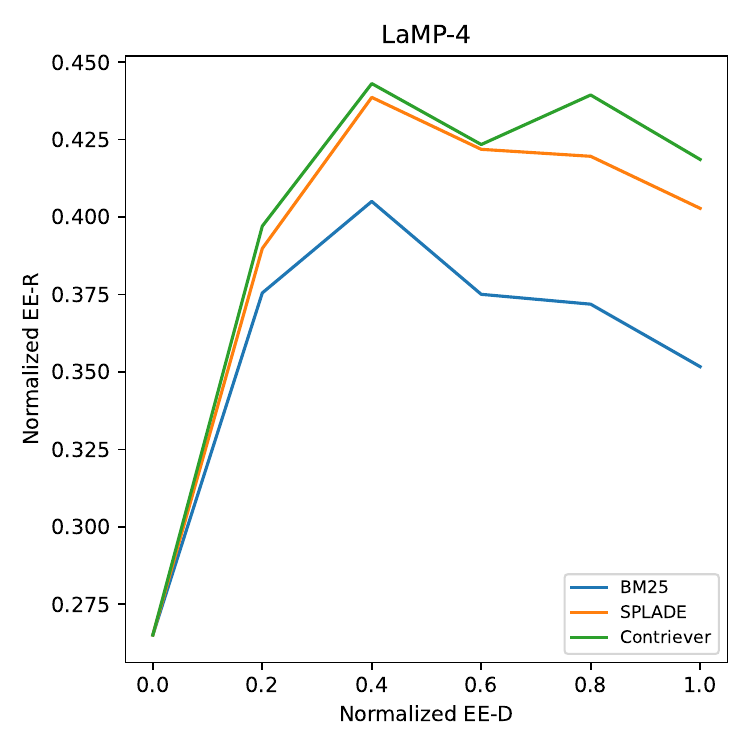}
        \caption{RetDisparity vs. Ranking Quality}
        \label{subfig:eed-eer-curve}
    \end{subfigure}
    \begin{subfigure}[b]{0.3\textwidth}
        \centering
        \includegraphics[width=\textwidth]{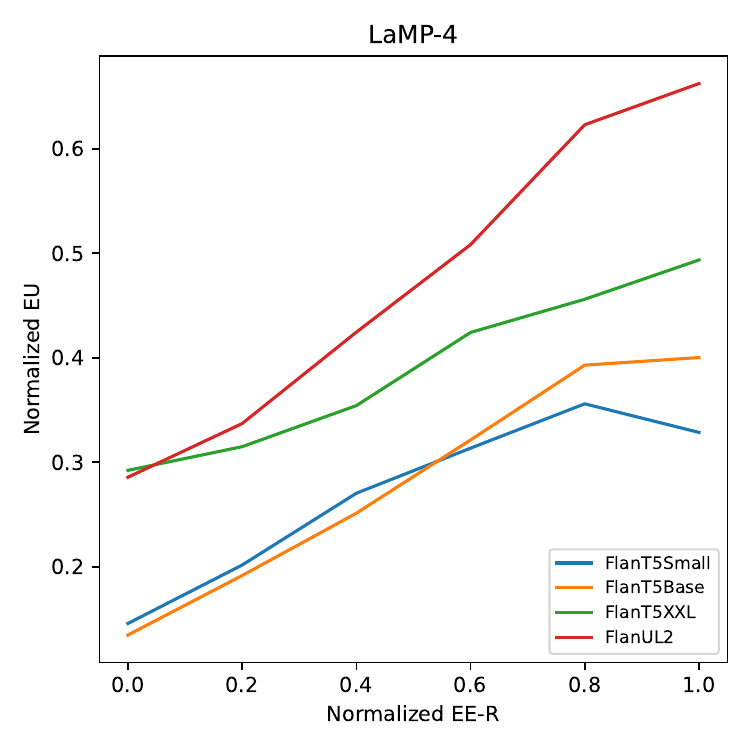}
        \caption{Ranking Quality vs. Utility}
        \label{subfig:eer-eu-curve}
    \end{subfigure}
    \begin{subfigure}[b]{0.3\textwidth}
        \centering
        \includegraphics[width=\textwidth]{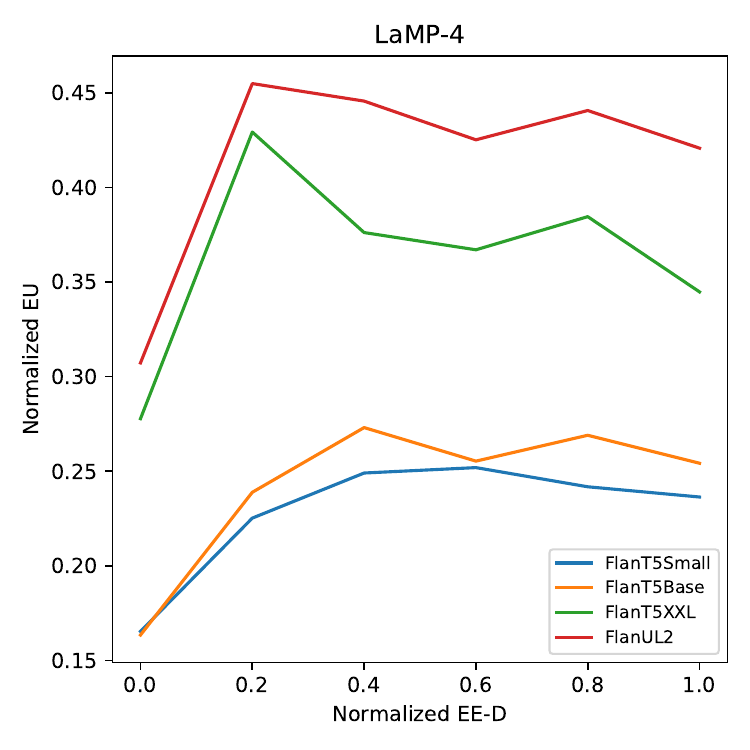}
        \caption{RetDisparity vs. Utility}
        \label{subfig:eed-eu-curve}
    \end{subfigure}
    \hfill
    \caption{Relationships between retrieval fairness, ranking quality, and generation quality. We empirically show that there is a positive relationship between ranking quality and utility, but trade-offs between retrieval fairness and ranking quality, as well as retrieval fairness and utility in RAG models. 
    The curves are plotted by interpolating the runs for each disparity interval and averaging across generators for \ref{subfig:eed-eer-curve} and across retrievers for \ref{subfig:eer-eu-curve} and \ref{subfig:eed-eu-curve}.
    RetDisparity refers to the disparity among retrieved items.
    }
    \label{fig:eed-eer-eu-set}
\end{figure*}

%% file: 07-discussion/tables/03-fairness-interval.tex
\begin{table*}[ht]
\centering
\begin{minipage}{\columnwidth}
\centering
\resizebox{\columnwidth}{!}{%
\begin{tabular}{cccccc}
\toprule
& \multicolumn{5}{c}{Disparity Intervals}\\
\cmidrule(lr){2-6} 
Model (baseline utility) & [0.0, 0.2) & [0.2, 0.4) & [0.4, 0.6) & [0.6, 0.8) & [0.8, 1.0) \\
\midrule
\multicolumn{6}{c}{\textbf{LaMP-1}} \\
\midrule
BM25+FlanT5Small (0.308) & -0.12 & -0.13 & -0.18 & -0.02 & -0.15 \\
BM25+FlanT5Base (0.670) & -0.20* & -0.04 & -0.08 & -0.05 & -0.02 \\
BM25+FlanT5XXL (0.531) & -0.07* & \cellcolor{green!25}+0.03 & \cellcolor{green!25}+0.02 & \cellcolor{green!25}+0.06 & \cellcolor{green!25}+0.11* \\
BM25+FlanUL2 (0.557) & -0.1* & -0.02 & -0.02 & \cellcolor{green!25}+0.01 & \cellcolor{green!25}+0.03 \\
\midrule
SPLADE+FlanT5Small (0.241) & -0.03 & -0.22 & \cellcolor{green!25}+0.19 & -0.04 & \cellcolor{green!25}+0.14 \\
SPLADE+FlanT5Base (0.646) & -0.15* & \cellcolor{green!25}+0.06 & \cellcolor{green!25}+0.08 & \cellcolor{green!25}0.00 & \cellcolor{green!25}+0.03 \\
SPLADE+FlanT5XXL (0.671) & -0.18* & -0.16 & \cellcolor{green!25}+0.05 & \cellcolor{green!25}+0.02 & \cellcolor{green!25}+0.01 \\ 
SPLADE+FlanUL2 (0.632) & -0.15* & -0.08 & \cellcolor{green!25}+0.02 & \cellcolor{green!25}+0.02 & \cellcolor{green!25}+0.01 \\ 
\midrule
Contriever+FlanT5Small (0.286) & -0.08 & -0.29 & -0.06 & \cellcolor{green!25}+0.03 & -0.14 \\
Contriever+FlanT5Base (0.637) & -0.16* & \cellcolor{green!25}+0.05 & -0.06 & \cellcolor{green!25}+0.03 & \cellcolor{green!25}0.00 \\
Contriever+FlanT5XXL (0.651) & -0.19* & -0.04 & -0.11 & \cellcolor{green!25}+0.03 & \cellcolor{green!25}0.00 \\
Contriever+FlanUL2 (0.620) & -0.16* & -0.09 & -0.05 & \cellcolor{green!25}+0.02 & -0.03 \\
\bottomrule
\end{tabular}
}
\end{minipage}
\begin{minipage}{\columnwidth}
\centering
\resizebox{\columnwidth}{!}{%
\begin{tabular}{cccccc}
\toprule
& \multicolumn{5}{c}{Disparity Intervals}\\
\cmidrule(lr){2-6} 
Model (baseline utility) & [0.0, 0.2) & [0.2, 0.4) & [0.4, 0.6) & [0.6, 0.8) & [0.8, 1.0) \\
\midrule
\multicolumn{6}{c}{\textbf{LaMP-4}} \\
\midrule
BM25+FlanT5Small (0.217) & -0.06* & \cellcolor{green!25}0.00 & \cellcolor{green!25}+0.02 & \cellcolor{green!25}+0.01 & \cellcolor{green!25}0.00 \\
BM25+FlanT5Base (0.223) & -0.06* & \cellcolor{green!25}0.00 & \cellcolor{green!25}+0.03 & \cellcolor{green!25}+0.01 & \cellcolor{green!25}+0.02 \\
BM25+FlanT5XXL (0.322) & -0.05* & \cellcolor{green!25}+0.11* & \cellcolor{green!25}+0.03 & \cellcolor{green!25}+0.03 & \cellcolor{green!25}+0.05* \\ 
BM25+FlanUL2 (0.381) & -0.07* & \cellcolor{green!25}+0.06* & \cellcolor{green!25}+0.07* & \cellcolor{green!25}+0.01 & \cellcolor{green!25}+0.04 \\
\midrule
SPLADE+FlanT5Small (0.235) & -0.07* & -0.01 & \cellcolor{green!25}+0.02 & \cellcolor{green!25}+0.03* & \cellcolor{green!25}+0.02 \\
SPLADE+FlanT5Base (0.268) & -0.10* & -0.03 & \cellcolor{green!25}+0.02 & \cellcolor{green!25}0.00 & \cellcolor{green!25}+0.02 \\
SPLADE+FlanT5XXL (0.342) & -0.06* & \cellcolor{green!25}+0.09* & \cellcolor{green!25}+0.05* & \cellcolor{green!25}+0.03 & \cellcolor{green!25}+0.04* \\ 
SPLADE+FlanUL2 (0.429) & -0.12* & \cellcolor{green!25}+0.04 & \cellcolor{green!25}+0.02 & \cellcolor{green!25}+0.01 & \cellcolor{green!25}+0.01 \\
\midrule
Contriever+FlanT5Small (0.254) & -0.09* & -0.02* & \cellcolor{green!25}0.00 & \cellcolor{green!25}+0.01 & \cellcolor{green!25}0.00 \\
Contriever+FlanT5Base (0.268) & -0.10* & -0.02 & \cellcolor{green!25}+0.01 & \cellcolor{green!25}0.00 & \cellcolor{green!25}+0.01 \\
Contriever+FlanT5XXL (0.367) & -0.09* & \cellcolor{green!25}+0.06* & \cellcolor{green!25}+0.01 & \cellcolor{green!25}+0.01 & \cellcolor{green!25}+0.03 \\
Contriever+FlanUL2 (0.449) & -0.15* & \cellcolor{green!25}+0.01 & -0.01 & \cellcolor{green!25}0.00 & \cellcolor{green!25}+0.02 \\
\bottomrule
\end{tabular}
}
\end{minipage}
\vspace{3pt}
\caption{Each value in the table is the difference between the utility of a baseline (deterministic) RAG model and the average utility of a fairer RAG model at a specific retrieval disparity interval. Nonnegative differences are highlighted. 
All LaMP tasks shows similar trend.
A superscript * denotes a statistically significant difference between the treatment performance and the baseline performance based on an unpaired Student's t-test ($p < 0.05$).
}
\label{tab:compare-baseline}
\vspace{-10pt}
\end{table*}

%% file: 07-discussion/figures/04-eae-lamp4.tex

\begin{figure*}[ht]
    \centering
    \begin{subfigure}[b]{0.3\textwidth}
        \centering
        \includegraphics[width=\textwidth]{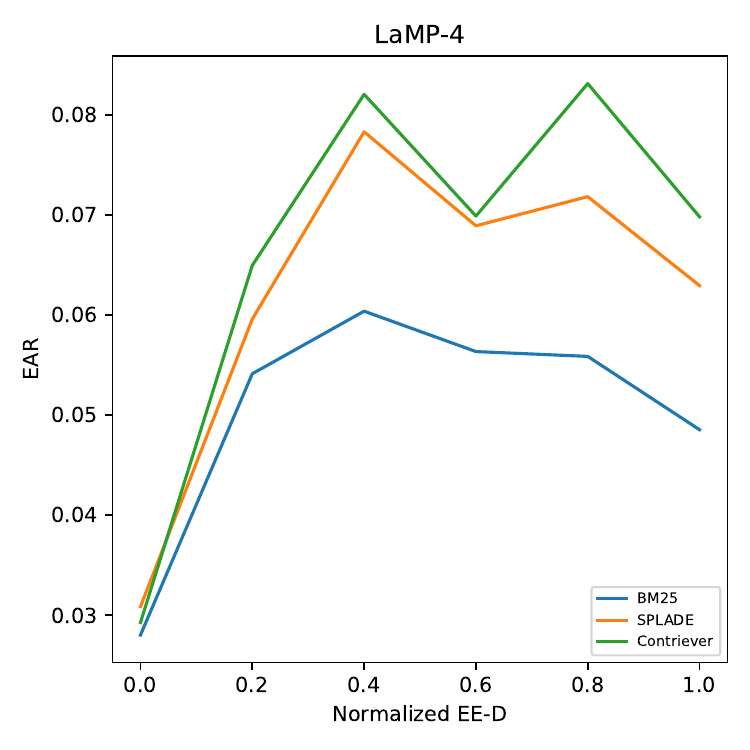}
        \caption{RetDisparity vs. Attribution Rate}
        \label{subfig:eed-ear-curve}
    \end{subfigure}
    \begin{subfigure}[b]{0.3\textwidth}
        \centering
        \includegraphics[width=\textwidth]{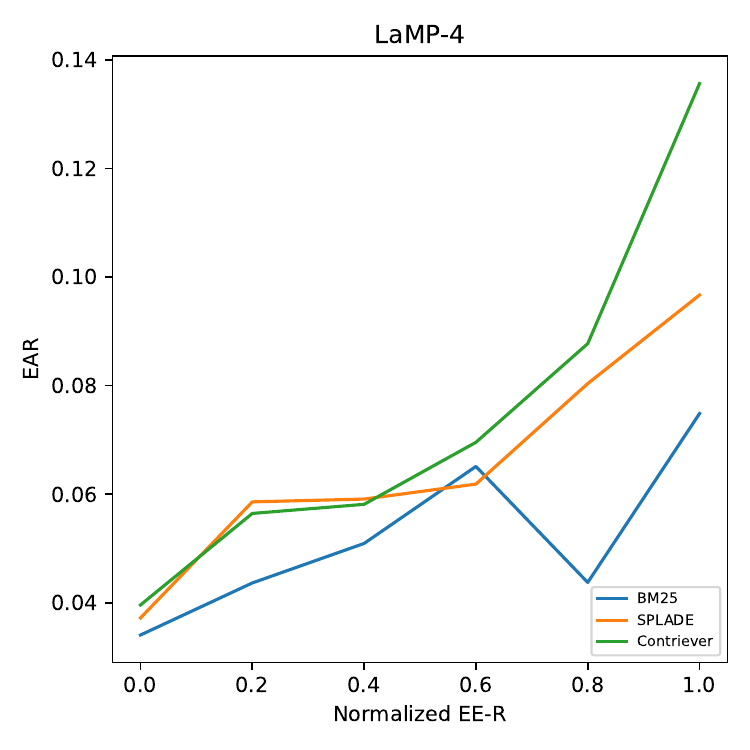}
        \caption{Ranking Quality vs. Attribution Rate}
        \label{subfig:eer-ear-curve}
    \end{subfigure}
    \begin{subfigure}[b]{0.3\textwidth}
        \centering
        \includegraphics[width=\textwidth]{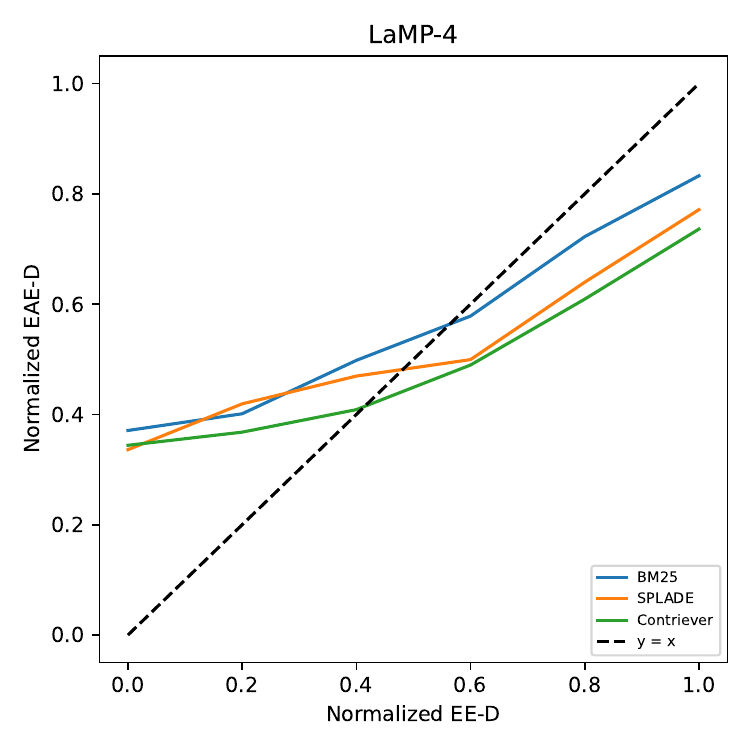}
        \caption{RetDisparity vs. AttDisparity}
        \label{subfig:eed-eaed-curve}
    \end{subfigure}
    \hfill
    \caption{
    Major relationships between retrieval fairness (RetDisparity), ranking quality, expected attribution rate, and consumption fairness (AttDisparity). We empirically show that there is a tradeoff between retrieval fairness and expected attribution rate as there is a positive relationship between ranking quality and attribution rate. Most importantly, we show that retrieval fairness does not necessarily directly propagate to the consumption fairness due to varying attribution rate of a generator. 
    The same interpolation methods are used as Figure \ref{fig:eed-eer-eu-set} and runs were averaged across generators. 
    RetDisparity refers to the disparity among retrieved items, while AttDisparity refers to the disparity among attributed sources.
    }
\end{figure*}

%% file: 08-discussion.tex

\section{Discussion}\label{sec:discussion}
\raggedbottom

\subsection{Higher System Utility with Fair Rankings}
Although there is a general trend of a fairness-utility tradeoff, we observe that certain levels of fairness can actually improve the utility of a baseline RAG model. Recent line of research have uncovered relevant findings: 1) generators are not robust to changes in the position of useful information \citep{liu2024lost}; 2) items with high retrieval scores often include distracting content that can reduce the system-effectiveness \citep{powerofnoise, ru2024ragchecker}; and 3) introducing some random documents can significantly boost the utility of RAG \citep{powerofnoise}.

Building on these existing results, we find that perturbing the initial ranking through stochastic sampling often can impact the performance of certain inference decisions and lead to changes in the system's expected end-performance. In our experiments, we observe that the expected utility generally increases within the fairness interval of [0.8, 1.0). This suggests that a fixed ranking from a deterministic ranker may be suboptimal for the generator, and that perturbing the ranking, along with the repositioning of items, not only improves expected end-performance but also enhances the fairness of the rankings.

Moreover, in fairness intervals where the system's expected utility improves, it is possible that either fewer distracting items were included in the ranking passed to the generator or useful, previously overlooked items (which may have been considered random) were introduced due to the ranking perturbation. However, while higher utility paired with increased item-fairness (even within fairness intervals as low as [0.4, 0.6)) may seem advantageous, practitioners should exercise caution. This could result in compensating providers of items irrelevant to user requests, particularly in scenarios where content providers are rewarded for contributing to inference outcomes.

\subsection{Attribution Rate and Consumption Fairness}
RAG practitioners should consider both the expected utility (EU) of the generated text and the expected attributed exposure disparity ($\overline{\text{EAE-D}}$) to ensure a balance between utility and fairness in the system's output. Achieving this balance requires careful adjustment of retrieval fairness to find an interval where the generated content remains both useful and fair in its exposure of sources. One important metric to monitor in this process is the attribution rate, which reflects the proportion of retrieved items that are actually cited or used in the final output. 

Monitoring expected attribution rate (EAR) provides valuable insights for understanding how fairness manifests in the final output, offering an essential complement to $\overline{\text{EAE-D}}$. For example, even if a ranking appears fair at the retrieval stage, a low attribution rate indicates that the generator ultimately utilizes only a small subset of the retrieved items, which can amplify disparities among the attributed sources. In such cases, fairness concerns at the generation stage may outweigh those at the retrieval stage. Conversely, when the generator attributes a larger portion of the retrieved set, disparities among the attributed sources may become less pronounced, as the system distributes attention more evenly across a broader set of items.

By comparing expected attribution rate with consumption fairness, practitioners can better disentangle the root causes of imbalances in the final output. Specifically, they can determine whether disparities stem from the retrieval stage, such as an inherently skewed ranking, or from the generator’s behavior in selectively focusing on only a few documents. This comparison can help diagnose fairness issues more effectively, enabling practitioners to refine both retrieval and generation components of a RAG system to achieve more equitable outcomes in the generated content.

\subsection{Measuring Consumption Fairness}
Ensuring consumption fairness while maintaining high utility of the generated text is a fundamental objective in building a fair RAG system. To achieve this, we measure the exposure of the final attributed items, as these are the items that human users will actually encounter in the system's output. This approach aligns with the real-world operation of RAG systems, particularly in applications like conversational QA or domain-specific assistants for general information or shopping, where the generated text is accompanied by explicitly cited documents or attributed items.  In such cases, measuring the fair exposure of the attributed items is critical, as consumption fairness can have a more direct impact on users compared to retrieval fairness, given that human users only consume what is explicitly surfaced by the system.

A key factor in accurately evaluating exposure-based retrieval fairness is the development of sophisticated machine-user browsing models. The work of \citet{liu2024lost} provides valuable insights for designing such advanced browsing models. In our experiments, we employed encoder-decoder models with a short retrieval context to get close to the assumption that equal attention is given to all top-$k$ retrieved items \cite{liu2024lost}. However, as the use of long-context becomes prevalent, accurately measuring exposure-based retrieval fairness becomes more challenging due to the generator’s tendency to allocate unequal attention to different items.

Despite these challenges, it is important to highlight that measuring consumption fairness is not dependent on machine-user browsing models. Unlike retrieval fairness, consumption fairness can be measured effectively in long-context models, regardless of how attentions are distributed across items by the generators. Therefore, focusing on consumption fairness ensures that the evaluation remains robust, even in scenarios where exposure patterns vary significantly across different models.


\subsection{Measurement of String Utility}
In line with the recent call for evaluating various valid output strings \citep{zhang2024diffusedistributionslanguage}, we recognize the need for a similar approach to better measure system utility across different rankings given. Recall that our experiments were designed to provide the generator with different rankings for the same query, leading to varied outputs. This approach is motivated by the idea that items not appearing in the top positions of deterministic rankings may still hold value and should be fairly considered by the system. In this context, the diverse outputs generated from different rankings may still be valid. However, we currently rely on a single target output string for comparison with predictions. Future work could focus on calculating the utility of diffuse predictions, enabling a more nuanced evaluation.

\subsection{Limitations}
We acknowledge that the evaluation cost of fair RAG systems can be high due to repeated sampling and inference steps. However, in production, only a single ranking is sampled, minimizing the impact on system latency. Also, a limitation in our utility labeling is that it considers single items, while multiple items may yield contrasting utility gains. Despite this, the strong correlation between ranking quality and system effectiveness suggests this approach reasonably approximates item-worthiness for evaluating the impact of fair ranking on RAG systems.

Another limitation lies in the ability to capture the exposure of items beyond those retrieved by the system. Since it is difficult to verify attribution across the entire corpus, our method focuses on measuring the exposure fairness of the attributed sources (e.g., those explicitly cited in the generated output). While this approach does not account for the broader set of potential exposures, it is useful for common RAG applications where the attributed sources are directly displayed to users. In such cases, ensuring fairness among these sources is critical, as they are likely to be the primary content users engage with.

%% file: 09-conclusion.tex
\section{Conclusion}
This study highlights the impact of fair rankings not only on the ranking and generation quality of RAG systems but also on the equitable attribution of sources in the final output. Through extensive analysis, we demonstrate that fairer RAG models can maintain—and in some cases even surpass—the generation quality of traditional approaches, challenging the assumption of an inherent tradeoff between fairness and effectiveness.
Our findings emphasize the importance of fair attribution, showing how improvements at the retrieval stage translate into more equitable exposure of the sources that appear in the generated text. By addressing disparities in both retrieval and attribution, we provide valuable insights for developing responsible and equitable RAG systems.

In future work, we hope to extend this framework to consider graded or missing judgments and exploring the different notions of fairness in RAG systems, ultimately advancing the field of trustworthy RAG systems research.
\footnote{We release our code and dataset at \url{https://github.com/kimdanny/Fair-RAG}.}

%% file: 97-appendix.tex
\section{Notation}\label{sec:app:notation}
\input{01-introduction/notation-table}

\section{Utility Labels Labeling Procedure}\label{sec:app:labeling-algo}
\begin{algorithm}[h]
  \caption{Labeling Procedure of Binary Utility Labels}\label{algo:utility-labels}
  \begin{algorithmic}[1]
    \State $\mathcal{D} = \{(\instance_1, \target_1),(\instance_2, \target_2),\cdots, (\instance_T, \target_T)\}$ \Comment{dataset of size $T$}
    \For{$(\instance_i, \target_i)\in\mathcal{D}$}
      \State $\stringUtility_i \gets \utilityMetric(\target_i, \mlmodel(\instance_i))$ \Comment{string utility of a baseline model without augmentation}
      \For{$\doc_j \in \collection$}
        \State $\hat{\target}_j \gets \mlmodel(\promptGenerator(\instance_i, \doc_j))$
        \State $\stringUtility_j \gets \utilityMetric(\target_i, \hat{\target}_j)$ \Comment{string utility of a generator augmented with one item}
        \State $\delta_j \gets (\stringUtility_j - \stringUtility_i)$ \Comment{marginal utility gained from the augmentation}
        \State $\worthiness{\doc_j}{\instance_i} \gets 0$
        \If {$\delta_j > 0$} \Comment{binary decision of item-worthiness by the utility-gain}
            \State $\worthiness{\doc_j}{\instance_i} \gets 1$
        \EndIf
      \EndFor
    \EndFor
  \end{algorithmic}
\end{algorithm}

\newpage
\section{Normalization of Metrics}\label{sec:app:normalization}

\subsection{Normalization of EE-D}\label{subsec:appendix-eed-norm}
The disparity measure EE-D should be normalized by its true upper and lower bound.
\begin{theorem}
$\ltwonorm{\expectedExposure}^{2} \in [0, \ltwonorm{\bar{\expectedExposure}}^{2}]$, where $\bar{\mathbf{\epsilon}}$ is an exposure vector derived from any deterministic ranking.
\end{theorem}
\begin{proof}\let\qed\relax
The lower bound is achieved by a uniform random policy. Each item $\doc$ will have exposure of $\frac{1}{n}$. However, it is reasonable to assume that it is approximately 0, since the size of most of the retrieval corpus is very large. Also, it is common that the corpus consists of majority of non-relevant items. The implication is that, for the optimal exposure, since the $n-m$ non-relevant items are shuffled amongst themselves, each will have an expected exposure of close to 0. Thus, assuming large $n$ and relatively small $m$,
\begin{equation}\label{eq:approx-zero}
    \frac{1}{n} < \frac{1}{n-m} \approx 0
\end{equation}
\noindent
For upper bound, recall that the $\expectedExposure$ is computed based on samples of rankings from a stochastic policy $\pi$. For relevance-independent browsing models (e.g., UM), all rankings $\rankedList\in S_n$ have identical exposure norms $\ltwonorm{\expectedExposure^{\rankedList}}^2$, where $\expectedExposure^{\rankedList}$ is the exposure of items for ranking sampled from $\pi$. Then,
\begin{align}
    \ltwonorm{\expectedExposure}^2 &= \ltwonorm{\mathbb{E}_{\rankedList\sim\pi} [\expectedExposure^{\rankedList}]}^2\\
    &\leq \mathbb{E}_{\rankedList\sim\pi}[\ltwonorm{\expectedExposure^{\rankedList}}^2]\\
    &= \mathbb{E}_{\rankedList\sim\pi}[\ltwonorm{\bar{\expectedExposure}}^2]
    = \ltwonorm{\bar{\expectedExposure}}^2
\end{align}
\end{proof}
\begin{corollary}
For machine user browsing model with \textit{top-k} consumption and equal attention to the top items, $\ltwonorm{\expectedExposure}^{2} \in [0, k]$
\end{corollary}
\noindent
With UM, the upper bound of EE-D, $\ltwonorm{\bar{\expectedExposure}}^{2}$ becomes $\sum_{i=1}^{n} {\text{UM}(i)}^2 = k$. Therefore, per query $q=\queryGenerator(\instance)$, we calculate a normalized EE-D
\begin{equation}
    \widetilde{\text{EE-D}}_\query = \ltwonorm{\expectedExposure}^{2} / k \quad \in [0, 1]
\end{equation}

\subsection{Normalization of EE-R}\label{subsec:appendix-eer-norm}
The relevance measure EE-R should be normalized by its true upper and lower bound.
\begin{theorem}
$\langle\expectedExposure, \targetExposure\rangle \in [0, \ltwonorm{\targetExposure}^{2}]$
\end{theorem}
\begin{proof}\let\qed\relax
The lower bound is achieved when $\expectedExposure$ becomes $\expectedExposure^-$, which is an exposure vector of the worst case deterministic ranking $\rankedList^-$ (permutations that rank all non-relevant items above relevant items). Given the assumption made from equation \ref{eq:approx-zero} and, $\collection^+$ and $\collection^-$, which are set of indices of relevant and non-relevant items, respectively,
\begin{align}
\langle\expectedExposure^-, \targetExposure\rangle 
&= \sum^{n}_{i=1} \expectedExposure^{-}_i \targetExposure_i\\
&\approx \sum_{i\in\collection^+} 0 \targetExposure_i + \sum_{i\in\collection^-} \expectedExposure^{-}_i 0 = 0 && \text{(from \ref{eq:approx-zero})}
\end{align}
\noindent
Intuitively, the upper bound is achieved when $\expectedExposure$ becomes $\targetExposure$, thus $\langle\targetExposure, \targetExposure\rangle = \ltwonorm{\targetExposure}^{2}$. 
Alternatively, we can show that any convex combination of optimal rankings will have a $\langle\expectedExposure, \targetExposure\rangle = \ltwonorm{\targetExposure}^{2}$.

Let $\targetExposure_\doc$ be the exposure of a relevant items, $S^*_n$ be the set of all optimal rankings, $w_{\rankedList'}$ be the weight on $\rankedList \in S^*_n$ such that $\sum_{\rankedList \in S^*_n} w_{\rankedList'} = 1$, and $\expectedExposure'$ be the exposure vector associated with $\rankedList'$.
\begin{align}
\langle\expectedExposure, \targetExposure\rangle 
&= \sum^{n}_{i=1} \expectedExposure_i \targetExposure_i = \sum_{i\in\collection^+} \expectedExposure_i \targetExposure_i &&\text{(from \ref{eq:approx-zero})}\\
&= \targetExposure_\doc \sum_{i\in\collection^+} \expectedExposure_i  && \text{(equal exposure principle)}\\
&= \targetExposure_\doc \sum_{i\in\collection^+} \sum_{\rankedList'\in S^*_n} w_{\rankedList'} \expectedExposure'_i\\
&= \targetExposure_\doc \sum_{\rankedList'\in S^*_n} w_{\rankedList'} \sum_{i\in\collection^+}  \expectedExposure'_i \\
& \leq \targetExposure_\doc \sum_{\rankedList'\in S^*_n} w_{\rankedList'} (m \targetExposure_\doc) && \text{(since $\expectedExposure'$ is optimal)}\\
&= \targetExposure_\doc (m \targetExposure_\doc) && \text{(since $\sum_{\rankedList \in S^*_n} w_{\rankedList'} = 1$)}\\
&= \sum_{i\in\collection^+} \targetExposure_i \targetExposure_i = \ltwonorm{\targetExposure}^{2}
\end{align}

\end{proof}
\begin{corollary}
For machine user browsing model with \textit{top-k} consumption and equal attention to the top items, the bound depends on $m$ and $k$. If $m \leq k$, $\langle\expectedExposure, \targetExposure\rangle \in [0, m + \frac{(k-m)^2}{n-m}]$. If $m>k$, $\langle\expectedExposure, \targetExposure\rangle \in [0, \frac{k^2}{m}]$
\end{corollary}
\noindent
With UM, the upper bound of EE-R can be calculated by the combination of equation \ref{eq:te-rel} and \ref{eq:te-non-rel}.\\ 
If $m \leq k$, $\ltwonorm{\targetExposure}^2$ becomes
\begin{align}
\sum_{i=\collection^+} 1^2 + \sum_{i=\collection^-}(\frac{k-m}{n-m})^2 = m + \frac{(k-m)^2}{n-m}
\end{align}
If $m > k$, $\ltwonorm{\targetExposure}^2$ becomes
\begin{align}
\sum_{i=\collection^+} (\frac{k}{m})^2 + \sum_{i=\collection^-} 0^2 = \frac{k^2}{m}
\end{align}
Therefore, depending on $m$ and $k$, per query $q=\queryGenerator(\instance)$, we calculate a normalized EE-R 
\begin{equation}
    \widetilde{\text{EE-R}}_\query = 
    \begin{cases}
    \langle\expectedExposure, \targetExposure\rangle / (m + \frac{(k-m)^2}{n-m}) & \text{($m \leq k$)}\\
    \langle\expectedExposure, \targetExposure\rangle / (\frac{k^2}{m}) & \text{($m > k$)}
    \end{cases}
    \in [0, 1]
\end{equation}

\subsection{Normalization of EU}\label{subsec:appendix-eu-norm}
Theoretically obtaining a true upper bound of a utility of a RAG model is challenging. Therefore, we approximate the true upper bound by the maximum of the empirically obtained utilities given a fixed RAG model $\text{RAG}(\rmodel, \mlmodel, k)$.

Recall that the string utility $\stringUtility_i = \utilityMetric(\target, \mlmodel(\promptGenerator(x, \rankedList_i)))$ is a utility with the sampled ranking $\rankedList_i \in \Sigma$.
We get a new set of utilities $\stringUtility^*_i = \utilityMetric(\target, \mlmodel(\promptGenerator(x, \rankedList^*_i)))$, where $\rankedList^*_i$ is an element of $\Sigma^*$, which is a set of sampled permutations (rankings) from a stochastic oracle retriever $\srmodel^*$. This stochastic oracle retriever is a policy that always places relevant items above non-relevant items, thus having $m!(n-m)!$ number of unique optimal permutations.

To get the empirical upper bound of the utility $\stringUtility_{max}$, we take the maximum of $\stringUtility$ and $\stringUtility^*$:
\begin{equation}
\stringUtility_{max} = max (\stringUtility_1, \stringUtility_2, \cdots, \stringUtility_N, \stringUtility^*_1, \stringUtility^*_2, \cdots, \stringUtility^*_N)
\end{equation}
\noindent
With $\stringUtility_{max}$, we \textit{max-normalize} EU.
Since $\frac{1}{N}\sum^{N}_{i=1} \frac{\stringUtility_i}{\stringUtility_{max}} $ is the same as $(\frac{1}{N}\sum^{N}_{i=1} \stringUtility_i) / \stringUtility_{max}$, per data instance $\instance$, we get a normalized EU
\begin{equation}
\widetilde{\text{EU}}_\instance = \frac{\text{EU}_\instance}{\stringUtility_{max}} \quad \in[0, 1]
\end{equation}

Normalization of EU is done to get the percentage of closeness to the optimal utility as all the utility values are scaled relative to the maximum value. In other words, the normalized EU value indicates how close the EU is to the maximum utility that the RAG system get get to. 

This is straightforward for \textit{higher-the-better} metrics, such as ROUGE and Accuracy. However, for \textit{lower-the-better} metrics such as MAE, we convert the scores to \textit{higher-the-better} by subtracting the scores from the true metric upper bound. This allows us to perform the same normalization operation and have the same interpretation of the normalized metric.

\onecolumn
\newpage
\newcolumntype{C}[1]{>{\centering\arraybackslash}p{#1}}

\section{Test Collection Statistics per Model}\label{app:data-stat}
\subsection{LaMP corpus statistics for \flantfivesmall}
\begin{table}[h]
\centering
\begin{tabular}{C{1.5cm} C{1.5cm} C{3cm} C{3cm} C{2.5cm}}
\hline
Corpus & \#queries & Avg \# Docs (Std) & Avg \# Useful Docs (Std) & Avg \% Useful Docs \\
\hline
LaMP-1 & 51 & 123.51 (82.66) & 9.08 (11.63) & 9.53 \\
LaMP-2 & 192 & 52.81 (46.21) & 7.98 (9.64) & 22.53 \\
LaMP-3 & 311 & 189.82 (134.33) & 65.88 (95.77) & 34.28 \\
LaMP-4 & 833 & 192.19 (195.28) & 40.72 (61.82) & 27.1 \\
LaMP-5 & 826 & 106.06 (71.47) & 26.18 (31.10) & 24.83 \\
LaMP-6 & 760 & 86.0 (52.66) & 27.78 (29.22) & 35.92 \\
LaMP-7 & 365 & 19.36 (18.4) & 8.23 (10.38) & 45.48 \\
\hline
\end{tabular}
\label{tab:filtered-lamp-stat-small}
\caption{LaMP corpus statistics for \flantfivesmall after filtering for fairness evaluation.}
\vspace{-10pt}
\end{table}

\subsection{LaMP corpus statistics for \flantfive}
\begin{table}[h]
\centering
\begin{tabular}{C{1.5cm} C{1.5cm} C{3cm} C{3cm} C{2.5cm}}
\hline
Corpus & \#queries & Avg \# Docs (Std) & Avg \# Useful Docs (Std) & Avg \% Useful Docs \\
\hline
LaMP-1 & 232 & 102.86 (61.88) & 20.07 (22.78) & 22.49 \\
LaMP-2 & 280 & 45.58 (42.0) & 10.45 (12.56) & 29.87 \\
LaMP-3 & 378 & 185.32 (128.43) & 73.82 (85.44) & 41.19 \\
LaMP-4 & 827 & 186.98 (193.52) & 49.79 (68.91) & 31.57 \\
LaMP-5 & 759 & 105.62 (69.56) & 26.09 (31.21) & 25.71 \\
LaMP-6 & 783 & 86.18 (52.97) & 30.11 (31.28) & 38.65 \\
LaMP-7 & 211 & 21.72 (16.09) & 6.96 (10.62) & 33.02 \\
\hline
\end{tabular}
\label{tab:filtered-lamp-stat-base}
\caption{LaMP corpus statistics for \flantfive after filtering for fairness evaluation.}
\vspace{-10pt}
\end{table}

\subsection{LaMP corpus statistics for \flantfivexxl}
\begin{table}[h]
\centering
\begin{tabular}{C{1.5cm} C{1.5cm} C{3cm} C{3cm} C{2.5cm}}
\hline
Corpus & \#queries & Avg \# Docs (Std) & Avg \# Useful Docs (Std) & Avg \% Useful Docs \\
\hline
LaMP-1 & 264 & 111.66 (69.45) & 25.12 (33.96) & 23.35 \\
LaMP-2 & 105 & 44.66 (42.82) & 11.32 (15.61) & 36.74 \\
LaMP-3 & 182 & 198.06 (151.09) & 41.86 (59.52) & 22.19 \\
LaMP-4 & 842 & 198.0 (200.82) & 54.07 (73.34) & 30.96 \\
LaMP-5 & 511 & 104.18 (68.73) & 23.1 (38.30) & 23.39 \\
LaMP-6 & 730 & 85.93 (52.46) & 34.89 (35.54) & 43.88 \\
LaMP-7 & 151 & 20.6 (16.39) & 8.58 (12.01) & 42.7 \\
\hline
\end{tabular}
\label{tab:filtered-lamp-stat-xxl}
\caption{LaMP corpus statistics for \flantfivexxl after filtering for fairness evaluation.}
\vspace{-10pt}
\end{table}

\subsection{LaMP corpus statistics for \flanultwo}
\begin{table}[h]
\centering
\begin{tabular}{C{1.5cm} C{1.5cm} C{3cm} C{3cm} C{2.5cm}}
\hline
Corpus & \#queries & Avg \# Docs (Std) & Avg \# Useful Docs (Std) & Avg \% Useful Docs \\
\hline
LaMP-1 & 411 & 111.0 (75.0) & 34.82 (42.54) & 31.52\\
LaMP-2 & 68 & 43.04 (38.87) & 12.94 (21.36) & 37.43\\
LaMP-3 & 140 & 206.95 (163.71) & 58.81 (89.21) & 30.34\\
LaMP-4 & 824 & 201.73 (205.19) & 54.23 (80.11) & 30.81\\
LaMP-5 & 459 & 105.67 (69.86) & 34.43 (55.53) & 32.35\\
LaMP-6 & 669 & 84.6 (51.67) & 36.16 (35.84) & 48.26\\
LaMP-7 & 245 & 19.22 (15.71) & 9.34 (10.93) & 51.17\\
\hline
\end{tabular}
\label{tab:filtered-lamp-stat-ul2}
\caption{LaMP corpus statistics for \flanultwo after filtering for fairness evaluation.}
\vspace{-10pt}
\end{table}







%% file: 01-introduction/notation-table.tex
\begin{table}[h]
    \centering
    \begin{tabular}{ll}\toprule
        Notation & Description \\\midrule
        $\instance$ & input instance \\
        $\target$ & output target \\
        $\queryGenerator(\instance)$ & query generation function\\
        $\query$ & query returned by $\queryGenerator(\instance)$\\
        $\doc$ & retrieval item (document) \\
        $\collection$ & stored retrievable items (corpus) \\
        $n$ & the number of $\doc$'s in $\collection$\\
        $m$ & the number of useful items in $\collection$\\
        \\
        $\rmodel(\query, \collection)$ & deterministic retriever \\
        $L$ & ranked list returned by $\rmodel(\query, \collection)$\\
        $\retrievalScores\in\mathbb{R}^n$ & retrieval scores returned by $\rmodel(\query, \collection)$ \\
        $N$ & sampling size for stochastic sampling\\
        $k$ & the number of $\doc$'s to retrieve per ranking\\
        $\srmodel(\retrievalScores, N, k)$ & stochastic ranking sampler \\
        $\rankedLists$ & set of $N$ sampled rankings returned by $\srmodel(\retrievalScores, N, k)$\\
        $\rankedList$ & sampled ranking $\in \rankedLists$ \\
        $\promptGenerator(\instance, \rankedList)$ & prompt generation function\\        
        $\xbar$ & prompt returned by $\promptGenerator(\instance, \rankedList)$\\
        $\mlmodel(\xbar)$ & language model \\
        $\hat{\target}$ & predicted output from $\mlmodel(\xbar)$ \\
        \\
        $\worthiness{\doc}{\instance}$ & worthiness of an item $\doc$ given an input $\instance$\\
        $\fairnessMetric(\rankedLists)$ & fairness metric of rankings\\ 
        $\relevanceMetric(\rankedLists)$ & relevance metric of rankings\\ 
        $\utilityMetric(\target, \hat{\target})$ & string utility metric\\
        $\expectedExposure\in\mathbb{R}^n$ & expected exposure of all items in $\collection$\\
        $\targetExposure\in\mathbb{R}^n$ & target exposure of all items in $\collection$\\
        \bottomrule
    \end{tabular}
    \caption{Notation.}
    \label{tab:notations}
    \vspace{-10pt}
\end{table}

%% file: 00-root.bbl

\begin{thebibliography}{63}


\ifx \showCODEN    \undefined \def \showCODEN     #1{\unskip}     \fi
\ifx \showDOI      \undefined \def \showDOI       #1{#1}\fi
\ifx \showISBNx    \undefined \def \showISBNx     #1{\unskip}     \fi
\ifx \showISBNxiii \undefined \def \showISBNxiii  #1{\unskip}     \fi
\ifx \showISSN     \undefined \def \showISSN      #1{\unskip}     \fi
\ifx \showLCCN     \undefined \def \showLCCN      #1{\unskip}     \fi
\ifx \shownote     \undefined \def \shownote      #1{#1}          \fi
\ifx \showarticletitle \undefined \def \showarticletitle #1{#1}   \fi
\ifx \showURL      \undefined \def \showURL       {\relax}        \fi
\providecommand\bibfield[2]{#2}
\providecommand\bibinfo[2]{#2}
\providecommand\natexlab[1]{#1}
\providecommand\showeprint[2][]{arXiv:#2}

\bibitem[Asai et~al\mbox{.}(2024)]%
        {asai2024selfrag}
\bibfield{author}{\bibinfo{person}{Akari Asai}, \bibinfo{person}{Zeqiu Wu}, \bibinfo{person}{Yizhong Wang}, \bibinfo{person}{Avirup Sil}, {and} \bibinfo{person}{Hannaneh Hajishirzi}.} \bibinfo{year}{2024}\natexlab{}.
\newblock \showarticletitle{Self-{RAG}: Learning to Retrieve, Generate, and Critique through Self-Reflection}. In \bibinfo{booktitle}{\emph{The Twelfth International Conference on Learning Representations}}.
\newblock


\bibitem[Bahri et~al\mbox{.}(2020)]%
        {Bahri2020ChoppyCT}
\bibfield{author}{\bibinfo{person}{Dara Bahri}, \bibinfo{person}{Yi Tay}, \bibinfo{person}{Che Zheng}, \bibinfo{person}{Donald Metzler}, {and} \bibinfo{person}{Andrew Tomkins}.} \bibinfo{year}{2020}\natexlab{}.
\newblock \showarticletitle{Choppy: Cut Transformer for Ranked List Truncation}.
\newblock \bibinfo{journal}{\emph{Proceedings of the 43rd International ACM SIGIR Conference on Research and Development in Information Retrieval}} (\bibinfo{year}{2020}).
\newblock


\bibitem[Balagopalan et~al\mbox{.}(2023)]%
        {Balagopalan23roleofrelevance}
\bibfield{author}{\bibinfo{person}{Aparna Balagopalan}, \bibinfo{person}{Abigail~Z. Jacobs}, {and} \bibinfo{person}{Asia~J. Biega}.} \bibinfo{year}{2023}\natexlab{}.
\newblock \showarticletitle{The Role of Relevance in Fair Ranking}. In \bibinfo{booktitle}{\emph{Proceedings of the 46th International ACM SIGIR Conference on Research and Development in Information Retrieval}} \emph{(\bibinfo{series}{SIGIR '23})}. \bibinfo{publisher}{Association for Computing Machinery}, \bibinfo{pages}{2650–2660}.
\newblock


\bibitem[Balan et~al\mbox{.}(2023)]%
        {balan:ekila}
\bibfield{author}{\bibinfo{person}{K. Balan}, \bibinfo{person}{S. Agarwal}, \bibinfo{person}{S. Jenni}, \bibinfo{person}{A. Parsons}, \bibinfo{person}{A. Gilbert}, {and} \bibinfo{person}{J. Collomosse}.} \bibinfo{year}{2023}\natexlab{}.
\newblock \showarticletitle{EKILA: Synthetic Media Provenance and Attribution for Generative Art}. In \bibinfo{booktitle}{\emph{2023 IEEE/CVF Conference on Computer Vision and Pattern Recognition Workshops (CVPRW)}}. \bibinfo{publisher}{IEEE Computer Society}, \bibinfo{pages}{913--922}.
\newblock


\bibitem[Biega et~al\mbox{.}(2018)]%
        {biega2018equity}
\bibfield{author}{\bibinfo{person}{Asia~J Biega}, \bibinfo{person}{Krishna~P Gummadi}, {and} \bibinfo{person}{Gerhard Weikum}.} \bibinfo{year}{2018}\natexlab{}.
\newblock \showarticletitle{Equity of attention: Amortizing individual fairness in rankings}. In \bibinfo{booktitle}{\emph{The 41st international acm sigir conference on research \& development in information retrieval}}. \bibinfo{pages}{405--414}.
\newblock


\bibitem[Bohnet et~al\mbox{.}(2022)]%
        {bohnet2022attributed}
\bibfield{author}{\bibinfo{person}{Bernd Bohnet}, \bibinfo{person}{Vinh~Q Tran}, \bibinfo{person}{Pat Verga}, \bibinfo{person}{Roee Aharoni}, \bibinfo{person}{Daniel Andor}, \bibinfo{person}{Livio~Baldini Soares}, \bibinfo{person}{Massimiliano Ciaramita}, \bibinfo{person}{Jacob Eisenstein}, \bibinfo{person}{Kuzman Ganchev}, \bibinfo{person}{Jonathan Herzig}, {et~al\mbox{.}}} \bibinfo{year}{2022}\natexlab{}.
\newblock \showarticletitle{Attributed question answering: Evaluation and modeling for attributed large language models}.
\newblock \bibinfo{journal}{\emph{arXiv preprint arXiv:2212.08037}} (\bibinfo{year}{2022}).
\newblock


\bibitem[Bruch et~al\mbox{.}(2020)]%
        {bruch_stochastic_LTR_2020}
\bibfield{author}{\bibinfo{person}{Sebastian Bruch}, \bibinfo{person}{Shuguang Han}, \bibinfo{person}{Michael Bendersky}, {and} \bibinfo{person}{Marc Najork}.} \bibinfo{year}{2020}\natexlab{}.
\newblock \showarticletitle{A {Stochastic} {Treatment} of {Learning} to {Rank} {Scoring} {Functions}}. In \bibinfo{booktitle}{\emph{Proceedings of the 13th {International} {Conference} on {Web} {Search} and {Data} {Mining}}} \emph{(\bibinfo{series}{{WSDM} '20})}. \bibinfo{publisher}{Association for Computing Machinery}, \bibinfo{pages}{61--69}.
\newblock


\bibitem[Caton and Haas(2020)]%
        {caton2020fairness-survey}
\bibfield{author}{\bibinfo{person}{Simon Caton} {and} \bibinfo{person}{Christian Haas}.} \bibinfo{year}{2020}\natexlab{}.
\newblock \showarticletitle{Fairness in machine learning: A survey}.
\newblock \bibinfo{journal}{\emph{Comput. Surveys}} (\bibinfo{year}{2020}).
\newblock


\bibitem[Chung et~al\mbox{.}(2024)]%
        {chung2022scaling-flant5}
\bibfield{author}{\bibinfo{person}{Hyung~Won Chung}, \bibinfo{person}{Le Hou}, \bibinfo{person}{Shayne Longpre}, \bibinfo{person}{Barret Zoph}, \bibinfo{person}{Yi Tay}, \bibinfo{person}{William Fedus}, \bibinfo{person}{Yunxuan Li}, \bibinfo{person}{Xuezhi Wang}, \bibinfo{person}{Mostafa Dehghani}, \bibinfo{person}{Siddhartha Brahma}, \bibinfo{person}{Albert Webson}, \bibinfo{person}{Shixiang~Shane Gu}, \bibinfo{person}{Zhuyun Dai}, \bibinfo{person}{Mirac Suzgun}, \bibinfo{person}{Xinyun Chen}, \bibinfo{person}{Aakanksha Chowdhery}, \bibinfo{person}{Alex Castro-Ros}, \bibinfo{person}{Marie Pellat}, \bibinfo{person}{Kevin Robinson}, \bibinfo{person}{Dasha Valter}, \bibinfo{person}{Sharan Narang}, \bibinfo{person}{Gaurav Mishra}, \bibinfo{person}{Adams Yu}, \bibinfo{person}{Vincent Zhao}, \bibinfo{person}{Yanping Huang}, \bibinfo{person}{Andrew Dai}, \bibinfo{person}{Hongkun Yu}, \bibinfo{person}{Slav Petrov}, \bibinfo{person}{Ed~H. Chi}, \bibinfo{person}{Jeff Dean}, \bibinfo{person}{Jacob Devlin},
  \bibinfo{person}{Adam Roberts}, \bibinfo{person}{Denny Zhou}, \bibinfo{person}{Quoc~V. Le}, {and} \bibinfo{person}{Jason Wei}.} \bibinfo{year}{2024}\natexlab{}.
\newblock \showarticletitle{Scaling Instruction-Finetuned Language Models}.
\newblock \bibinfo{journal}{\emph{Journal of Machine Learning Research}} \bibinfo{volume}{25}, \bibinfo{number}{70} (\bibinfo{year}{2024}), \bibinfo{pages}{1--53}.
\newblock
\urldef\tempurl%
\url{http://jmlr.org/papers/v25/23-0870.html}
\showURL{%
\tempurl}


\bibitem[Cuconasu et~al\mbox{.}(2024)]%
        {powerofnoise}
\bibfield{author}{\bibinfo{person}{Florin Cuconasu}, \bibinfo{person}{Giovanni Trappolini}, \bibinfo{person}{Federico Siciliano}, \bibinfo{person}{Simone Filice}, \bibinfo{person}{Cesare Campagnano}, \bibinfo{person}{Yoelle Maarek}, \bibinfo{person}{Nicola Tonellotto}, {and} \bibinfo{person}{Fabrizio Silvestri}.} \bibinfo{year}{2024}\natexlab{}.
\newblock \showarticletitle{The Power of Noise: Redefining Retrieval for RAG Systems}. In \bibinfo{booktitle}{\emph{Proceedings of the 47th International ACM SIGIR Conference on Research and Development in Information Retrieval}} \emph{(\bibinfo{series}{SIGIR '24})}. \bibinfo{publisher}{Association for Computing Machinery}, \bibinfo{pages}{719–729}.
\newblock


\bibitem[Diaz et~al\mbox{.}(2020)]%
        {diaz_2020_stochastic_ranking}
\bibfield{author}{\bibinfo{person}{Fernando Diaz}, \bibinfo{person}{Bhaskar Mitra}, \bibinfo{person}{Michael~D. Ekstrand}, \bibinfo{person}{Asia~J. Biega}, {and} \bibinfo{person}{Ben Carterette}.} \bibinfo{year}{2020}\natexlab{}.
\newblock \showarticletitle{Evaluating {Stochastic} {Rankings} with {Expected} {Exposure}}. In \bibinfo{booktitle}{\emph{Proceedings of the 29th {ACM} {International} {Conference} on {Information} \& {Knowledge} {Management}}} \emph{(\bibinfo{series}{{CIKM} '20})}. \bibinfo{publisher}{Association for Computing Machinery}, \bibinfo{pages}{275--284}.
\newblock


\bibitem[Ekstrand et~al\mbox{.}(2024)]%
        {ekstrand24notjustalgo}
\bibfield{author}{\bibinfo{person}{Michael~D. Ekstrand}, \bibinfo{person}{Lex Beattie}, \bibinfo{person}{Maria~Soledad Pera}, {and} \bibinfo{person}{Henriette Cramer}.} \bibinfo{year}{2024}\natexlab{}.
\newblock \showarticletitle{Not Just Algorithms: Strategically Addressing Consumer Impacts in Information Retrieval}. In \bibinfo{booktitle}{\emph{Advances in Information Retrieval}}. \bibinfo{publisher}{Springer Nature Switzerland}, \bibinfo{pages}{314--335}.
\newblock


\bibitem[Ekstrand et~al\mbox{.}(2022)]%
        {ekstrand-diaz2022fairness-survey}
\bibfield{author}{\bibinfo{person}{Michael~D Ekstrand}, \bibinfo{person}{Anubrata Das}, \bibinfo{person}{Robin Burke}, {and} \bibinfo{person}{Fernando Diaz}.} \bibinfo{year}{2022}\natexlab{}.
\newblock \showarticletitle{Fairness in information access systems}.
\newblock \bibinfo{journal}{\emph{Foundations and Trends{\textregistered} in Information Retrieval}} \bibinfo{volume}{16}, \bibinfo{number}{1-2} (\bibinfo{year}{2022}), \bibinfo{pages}{1--177}.
\newblock


\bibitem[Es et~al\mbox{.}(2024)]%
        {es-etal-2024-ragas}
\bibfield{author}{\bibinfo{person}{Shahul Es}, \bibinfo{person}{Jithin James}, \bibinfo{person}{Luis Espinosa~Anke}, {and} \bibinfo{person}{Steven Schockaert}.} \bibinfo{year}{2024}\natexlab{}.
\newblock \showarticletitle{{RAGA}s: Automated Evaluation of Retrieval Augmented Generation}. In \bibinfo{booktitle}{\emph{Proceedings of the 18th Conference of the European Chapter of the Association for Computational Linguistics: System Demonstrations}}, \bibfield{editor}{\bibinfo{person}{Nikolaos Aletras} {and} \bibinfo{person}{Orphee De~Clercq}} (Eds.). \bibinfo{publisher}{Association for Computational Linguistics}, \bibinfo{pages}{150--158}.
\newblock


\bibitem[Formal et~al\mbox{.}(2022)]%
        {splade}
\bibfield{author}{\bibinfo{person}{Thibault Formal}, \bibinfo{person}{Carlos Lassance}, \bibinfo{person}{Benjamin Piwowarski}, {and} \bibinfo{person}{St\'{e}phane Clinchant}.} \bibinfo{year}{2022}\natexlab{}.
\newblock \showarticletitle{From Distillation to Hard Negative Sampling: Making Sparse Neural IR Models More Effective}. In \bibinfo{booktitle}{\emph{Proceedings of the 45th International ACM SIGIR Conference on Research and Development in Information Retrieval}} (Madrid, Spain) \emph{(\bibinfo{series}{SIGIR '22})}. \bibinfo{publisher}{Association for Computing Machinery}, \bibinfo{address}{New York, NY, USA}, \bibinfo{pages}{2353–2359}.
\newblock
\showISBNx{9781450387323}
\urldef\tempurl%
\url{https://doi.org/10.1145/3477495.3531857}
\showDOI{\tempurl}


\bibitem[Gao et~al\mbox{.}(2023a)]%
        {gao:rarr}
\bibfield{author}{\bibinfo{person}{Luyu Gao}, \bibinfo{person}{Zhuyun Dai}, \bibinfo{person}{Panupong Pasupat}, \bibinfo{person}{Anthony Chen}, \bibinfo{person}{Arun~Tejasvi Chaganty}, \bibinfo{person}{Yicheng Fan}, \bibinfo{person}{Vincent Zhao}, \bibinfo{person}{Ni Lao}, \bibinfo{person}{Hongrae Lee}, \bibinfo{person}{Da-Cheng Juan}, {and} \bibinfo{person}{Kelvin Guu}.} \bibinfo{year}{2023}\natexlab{a}.
\newblock \showarticletitle{{RARR}: Researching and Revising What Language Models Say, Using Language Models}. In \bibinfo{booktitle}{\emph{Proceedings of the 61st Annual Meeting of the Association for Computational Linguistics (Volume 1: Long Papers)}}. \bibinfo{publisher}{Association for Computational Linguistics}, \bibinfo{pages}{16477--16508}.
\newblock


\bibitem[Gao et~al\mbox{.}(2023b)]%
        {gao:alce}
\bibfield{author}{\bibinfo{person}{Tianyu Gao}, \bibinfo{person}{Howard Yen}, \bibinfo{person}{Jiatong Yu}, {and} \bibinfo{person}{Danqi Chen}.} \bibinfo{year}{2023}\natexlab{b}.
\newblock \showarticletitle{Enabling Large Language Models to Generate Text with Citations}. In \bibinfo{booktitle}{\emph{Proceedings of the 2023 Conference on Empirical Methods in Natural Language Processing}}. \bibinfo{publisher}{Association for Computational Linguistics}, \bibinfo{pages}{6465--6488}.
\newblock
\urldef\tempurl%
\url{https://doi.org/10.18653/v1/2023.emnlp-main.398}
\showDOI{\tempurl}


\bibitem[Guiver and Snelson(2009)]%
        {guiver_bayesian_PL_2009}
\bibfield{author}{\bibinfo{person}{John Guiver} {and} \bibinfo{person}{Edward Snelson}.} \bibinfo{year}{2009}\natexlab{}.
\newblock \showarticletitle{Bayesian inference for {Plackett}-{Luce} ranking models}. In \bibinfo{booktitle}{\emph{Proceedings of the 26th {Annual} {International} {Conference} on {Machine} {Learning}}}. \bibinfo{publisher}{ACM}, \bibinfo{pages}{377--384}.
\newblock


\bibitem[Gumbel(1954)]%
        {gumbel1954statistical}
\bibfield{author}{\bibinfo{person}{Emil~Julius Gumbel}.} \bibinfo{year}{1954}\natexlab{}.
\newblock \bibinfo{booktitle}{\emph{Statistical theory of extreme values and some practical applications: a series of lectures}}. Vol.~\bibinfo{volume}{33}.
\newblock \bibinfo{publisher}{US Government Printing Office}.
\newblock


\bibitem[Guu et~al\mbox{.}(2020)]%
        {guu-realm}
\bibfield{author}{\bibinfo{person}{Kelvin Guu}, \bibinfo{person}{Kenton Lee}, \bibinfo{person}{Zora Tung}, \bibinfo{person}{Panupong Pasupat}, {and} \bibinfo{person}{Ming-Wei Chang}.} \bibinfo{year}{2020}\natexlab{}.
\newblock \showarticletitle{REALM: Retrieval-Augmented Language Model Pre-Training}. In \bibinfo{booktitle}{\emph{Proceedings of the 37th International Conference on Machine Learning}} \emph{(\bibinfo{series}{ICML'20})}. \bibinfo{publisher}{JMLR.org}, Article \bibinfo{articleno}{368}, \bibinfo{numpages}{10}~pages.
\newblock


\bibitem[He et~al\mbox{.}(2024)]%
        {he-etal-2024-never}
\bibfield{author}{\bibinfo{person}{Junqing He}, \bibinfo{person}{Kunhao Pan}, \bibinfo{person}{Xiaoqun Dong}, \bibinfo{person}{Zhuoyang Song}, \bibinfo{person}{LiuYiBo LiuYiBo}, \bibinfo{person}{Qianguosun Qianguosun}, \bibinfo{person}{Yuxin Liang}, \bibinfo{person}{Hao Wang}, \bibinfo{person}{Enming Zhang}, {and} \bibinfo{person}{Jiaxing Zhang}.} \bibinfo{year}{2024}\natexlab{}.
\newblock \showarticletitle{Never Lost in the Middle: Mastering Long-Context Question Answering with Position-Agnostic Decompositional Training}. In \bibinfo{booktitle}{\emph{Proceedings of the 62nd Annual Meeting of the Association for Computational Linguistics (Volume 1: Long Papers)}}, \bibfield{editor}{\bibinfo{person}{Lun-Wei Ku}, \bibinfo{person}{Andre Martins}, {and} \bibinfo{person}{Vivek Srikumar}} (Eds.). \bibinfo{publisher}{Association for Computational Linguistics}, \bibinfo{address}{Bangkok, Thailand}, \bibinfo{pages}{13628--13642}.
\newblock
\urldef\tempurl%
\url{https://doi.org/10.18653/v1/2024.acl-long.736}
\showDOI{\tempurl}


\bibitem[Henderson et~al\mbox{.}(2023)]%
        {henderson:fm-fair-use}
\bibfield{author}{\bibinfo{person}{Peter Henderson}, \bibinfo{person}{Xuechen Li}, \bibinfo{person}{Dan Jurafsky}, \bibinfo{person}{Tatsunori Hashimoto}, \bibinfo{person}{Mark~A. Lemley}, {and} \bibinfo{person}{Percy Liang}.} \bibinfo{year}{2023}\natexlab{}.
\newblock \showarticletitle{Foundation Models and Fair Use}.
\newblock \bibinfo{journal}{\emph{Journal of Machine Learning Research}} \bibinfo{volume}{24}, \bibinfo{number}{400} (\bibinfo{year}{2023}), \bibinfo{pages}{1--79}.
\newblock


\bibitem[Hofst{\"a}tter et~al\mbox{.}(2023)]%
        {FiD-Light}
\bibfield{author}{\bibinfo{person}{Sebastian Hofst{\"a}tter}, \bibinfo{person}{Jiecao Chen}, \bibinfo{person}{Karthik Raman}, {and} \bibinfo{person}{Hamed Zamani}.} \bibinfo{year}{2023}\natexlab{}.
\newblock \showarticletitle{Fid-light: Efficient and effective retrieval-augmented text generation}. In \bibinfo{booktitle}{\emph{Proceedings of the 46th International ACM SIGIR Conference on Research and Development in Information Retrieval}}. \bibinfo{pages}{1437--1447}.
\newblock


\bibitem[Honovich et~al\mbox{.}(2022)]%
        {honovich-2022-true}
\bibfield{author}{\bibinfo{person}{Or Honovich}, \bibinfo{person}{Roee Aharoni}, \bibinfo{person}{Jonathan Herzig}, \bibinfo{person}{Hagai Taitelbaum}, \bibinfo{person}{Doron Kukliansy}, \bibinfo{person}{Vered Cohen}, \bibinfo{person}{Thomas Scialom}, \bibinfo{person}{Idan Szpektor}, \bibinfo{person}{Avinatan Hassidim}, {and} \bibinfo{person}{Yossi Matias}.} \bibinfo{year}{2022}\natexlab{}.
\newblock \showarticletitle{{TRUE}: Re-evaluating Factual Consistency Evaluation}. In \bibinfo{booktitle}{\emph{Proceedings of the 2022 Conference of the North American Chapter of the Association for Computational Linguistics: Human Language Technologies}}, \bibfield{editor}{\bibinfo{person}{Marine Carpuat}, \bibinfo{person}{Marie-Catherine de~Marneffe}, {and} \bibinfo{person}{Ivan~Vladimir Meza~Ruiz}} (Eds.). \bibinfo{publisher}{Association for Computational Linguistics}, \bibinfo{address}{Seattle, United States}, \bibinfo{pages}{3905--3920}.
\newblock
\urldef\tempurl%
\url{https://doi.org/10.18653/v1/2022.naacl-main.287}
\showDOI{\tempurl}


\bibitem[Hsieh et~al\mbox{.}(2024)]%
        {hsieh-etal-2024-found}
\bibfield{author}{\bibinfo{person}{Cheng-Yu Hsieh}, \bibinfo{person}{Yung-Sung Chuang}, \bibinfo{person}{Chun-Liang Li}, \bibinfo{person}{Zifeng Wang}, \bibinfo{person}{Long Le}, \bibinfo{person}{Abhishek Kumar}, \bibinfo{person}{James Glass}, \bibinfo{person}{Alexander Ratner}, \bibinfo{person}{Chen-Yu Lee}, \bibinfo{person}{Ranjay Krishna}, {and} \bibinfo{person}{Tomas Pfister}.} \bibinfo{year}{2024}\natexlab{}.
\newblock \showarticletitle{Found in the middle: Calibrating Positional Attention Bias Improves Long Context Utilization}. In \bibinfo{booktitle}{\emph{Findings of the Association for Computational Linguistics: ACL 2024}}, \bibfield{editor}{\bibinfo{person}{Lun-Wei Ku}, \bibinfo{person}{Andre Martins}, {and} \bibinfo{person}{Vivek Srikumar}} (Eds.). \bibinfo{publisher}{Association for Computational Linguistics}, \bibinfo{address}{Bangkok, Thailand}, \bibinfo{pages}{14982--14995}.
\newblock
\urldef\tempurl%
\url{https://doi.org/10.18653/v1/2024.findings-acl.890}
\showDOI{\tempurl}


\bibitem[Izacard et~al\mbox{.}(2022)]%
        {izacard2022contriever}
\bibfield{author}{\bibinfo{person}{Gautier Izacard}, \bibinfo{person}{Mathilde Caron}, \bibinfo{person}{Lucas Hosseini}, \bibinfo{person}{Sebastian Riedel}, \bibinfo{person}{Piotr Bojanowski}, \bibinfo{person}{Armand Joulin}, {and} \bibinfo{person}{Edouard Grave}.} \bibinfo{year}{2022}\natexlab{}.
\newblock \showarticletitle{Unsupervised Dense Information Retrieval with Contrastive Learning}.
\newblock \bibinfo{journal}{\emph{Transactions on Machine Learning Research}} (\bibinfo{year}{2022}).
\newblock
\showISSN{2835-8856}


\bibitem[Izacard et~al\mbox{.}(2023)]%
        {izacard_few-shot_2022}
\bibfield{author}{\bibinfo{person}{Gautier Izacard}, \bibinfo{person}{Patrick Lewis}, \bibinfo{person}{Maria Lomeli}, \bibinfo{person}{Lucas Hosseini}, \bibinfo{person}{Fabio Petroni}, \bibinfo{person}{Timo Schick}, \bibinfo{person}{Jane Dwivedi-Yu}, \bibinfo{person}{Armand Joulin}, \bibinfo{person}{Sebastian Riedel}, {and} \bibinfo{person}{Edouard Grave}.} \bibinfo{year}{2023}\natexlab{}.
\newblock \showarticletitle{Atlas: Few-shot learning with retrieval augmented language models}.
\newblock \bibinfo{journal}{\emph{Journal of Machine Learning Research}} \bibinfo{volume}{24}, \bibinfo{number}{251} (\bibinfo{year}{2023}), \bibinfo{pages}{1--43}.
\newblock


\bibitem[Jaenich et~al\mbox{.}(2024)]%
        {JaenichSigir24FairReranking}
\bibfield{author}{\bibinfo{person}{Thomas Jaenich}, \bibinfo{person}{Graham McDonald}, {and} \bibinfo{person}{Iadh Ounis}.} \bibinfo{year}{2024}\natexlab{}.
\newblock \showarticletitle{Fairness-Aware Exposure Allocation via Adaptive Reranking}. In \bibinfo{booktitle}{\emph{Proceedings of the 47th International ACM SIGIR Conference on Research and Development in Information Retrieval}} \emph{(\bibinfo{series}{SIGIR '24})}. \bibinfo{publisher}{Association for Computing Machinery}, \bibinfo{pages}{1504–1513}.
\newblock


\bibitem[Jiang et~al\mbox{.}(2024)]%
        {Jiang24ItemsideLLM}
\bibfield{author}{\bibinfo{person}{Meng Jiang}, \bibinfo{person}{Keqin Bao}, \bibinfo{person}{Jizhi Zhang}, \bibinfo{person}{Wenjie Wang}, \bibinfo{person}{Zhengyi Yang}, \bibinfo{person}{Fuli Feng}, {and} \bibinfo{person}{Xiangnan He}.} \bibinfo{year}{2024}\natexlab{}.
\newblock \showarticletitle{Item-side Fairness of Large Language Model-based Recommendation System}. In \bibinfo{booktitle}{\emph{Proceedings of the ACM on Web Conference 2024}} \emph{(\bibinfo{series}{WWW '24})}. \bibinfo{publisher}{Association for Computing Machinery}, \bibinfo{pages}{4717–4726}.
\newblock


\bibitem[Kelly et~al\mbox{.}(2013)]%
        {nsf-workshop-human-ir}
\bibfield{author}{\bibinfo{person}{Diane Kelly}, \bibinfo{person}{Jaime Arguello}, {and} \bibinfo{person}{Robert Capra}.} \bibinfo{year}{2013}\natexlab{}.
\newblock \showarticletitle{NSF workshop on task-based information search systems}.
\newblock \bibinfo{journal}{\emph{SIGIR Forum}} \bibinfo{volume}{47}, \bibinfo{number}{2} (\bibinfo{year}{2013}), \bibinfo{pages}{116–127}.
\newblock
\showISSN{0163-5840}


\bibitem[Khandelwal et~al\mbox{.}(2020)]%
        {Khandelwal2020Generalization}
\bibfield{author}{\bibinfo{person}{Urvashi Khandelwal}, \bibinfo{person}{Omer Levy}, \bibinfo{person}{Dan Jurafsky}, \bibinfo{person}{Luke Zettlemoyer}, {and} \bibinfo{person}{Mike Lewis}.} \bibinfo{year}{2020}\natexlab{}.
\newblock \showarticletitle{Generalization through Memorization: Nearest Neighbor Language Models}. In \bibinfo{booktitle}{\emph{International Conference on Learning Representations}}.
\newblock


\bibitem[Kim et~al\mbox{.}(2024)]%
        {kim2024reml}
\bibfield{author}{\bibinfo{person}{To~Eun Kim}, \bibinfo{person}{Alireza Salemi}, \bibinfo{person}{Andrew Drozdov}, \bibinfo{person}{Fernando Diaz}, {and} \bibinfo{person}{Hamed Zamani}.} \bibinfo{year}{2024}\natexlab{}.
\newblock \showarticletitle{Retrieval-Enhanced Machine Learning: Synthesis and Opportunities}.
\newblock \bibinfo{journal}{\emph{arXiv preprint arXiv:2407.12982}} (\bibinfo{year}{2024}).
\newblock


\bibitem[Lewis et~al\mbox{.}(2020)]%
        {Lewis+al:2020}
\bibfield{author}{\bibinfo{person}{Patrick S.~H. Lewis}, \bibinfo{person}{Ethan Perez}, \bibinfo{person}{Aleksandra Piktus}, \bibinfo{person}{Fabio Petroni}, \bibinfo{person}{Vladimir Karpukhin}, \bibinfo{person}{Naman Goyal}, \bibinfo{person}{Heinrich K{\"{u}}ttler}, \bibinfo{person}{Mike Lewis}, \bibinfo{person}{Wen{-}tau Yih}, \bibinfo{person}{Tim Rockt{\"{a}}schel}, \bibinfo{person}{Sebastian Riedel}, {and} \bibinfo{person}{Douwe Kiela}.} \bibinfo{year}{2020}\natexlab{}.
\newblock \showarticletitle{Retrieval-Augmented Generation for Knowledge-Intensive {NLP} Tasks}. In \bibinfo{booktitle}{\emph{Advances in Neural Information Processing Systems 33: Annual Conference on Neural Information Processing Systems 2020, NeurIPS 2020, December 6-12, 2020, virtual}}, \bibfield{editor}{\bibinfo{person}{Hugo Larochelle}, \bibinfo{person}{Marc'Aurelio Ranzato}, \bibinfo{person}{Raia Hadsell}, \bibinfo{person}{Maria{-}Florina Balcan}, {and} \bibinfo{person}{Hsuan{-}Tien Lin}} (Eds.).
\newblock


\bibitem[Liu et~al\mbox{.}(2024)]%
        {liu2024lost}
\bibfield{author}{\bibinfo{person}{Nelson~F Liu}, \bibinfo{person}{Kevin Lin}, \bibinfo{person}{John Hewitt}, \bibinfo{person}{Ashwin Paranjape}, \bibinfo{person}{Michele Bevilacqua}, \bibinfo{person}{Fabio Petroni}, {and} \bibinfo{person}{Percy Liang}.} \bibinfo{year}{2024}\natexlab{}.
\newblock \showarticletitle{Lost in the middle: How language models use long contexts}.
\newblock \bibinfo{journal}{\emph{Transactions of the Association for Computational Linguistics}}  \bibinfo{volume}{12} (\bibinfo{year}{2024}), \bibinfo{pages}{157--173}.
\newblock


\bibitem[Liu(2019)]%
        {liu2019roberta}
\bibfield{author}{\bibinfo{person}{Yinhan Liu}.} \bibinfo{year}{2019}\natexlab{}.
\newblock \showarticletitle{Roberta: A robustly optimized bert pretraining approach}.
\newblock \bibinfo{journal}{\emph{arXiv preprint arXiv:1907.11692}}  \bibinfo{volume}{364} (\bibinfo{year}{2019}).
\newblock


\bibitem[Lyu et~al\mbox{.}(2023)]%
        {chen:resp-ai-gen}
\bibfield{author}{\bibinfo{person}{Lingjuan Lyu}, \bibinfo{person}{C Chen}, {and} \bibinfo{person}{J Fu}.} \bibinfo{year}{2023}\natexlab{}.
\newblock \showarticletitle{A Pathway Towards Responsible AI Generated Content.}. In \bibinfo{booktitle}{\emph{IJCAI}}. \bibinfo{pages}{7033--7038}.
\newblock


\bibitem[Maddison et~al\mbox{.}(2017)]%
        {maddison2017gumbel}
\bibfield{author}{\bibinfo{person}{Chris~J. Maddison}, \bibinfo{person}{Andriy Mnih}, {and} \bibinfo{person}{Yee~Whye Teh}.} \bibinfo{year}{2017}\natexlab{}.
\newblock \showarticletitle{The Concrete Distribution: A Continuous Relaxation of Discrete Random Variables}. In \bibinfo{booktitle}{\emph{International Conference on Learning Representations}}.
\newblock


\bibitem[Mehrotra et~al\mbox{.}(2017)]%
        {Mehrotra17auditing}
\bibfield{author}{\bibinfo{person}{Rishabh Mehrotra}, \bibinfo{person}{Ashton Anderson}, \bibinfo{person}{Fernando Diaz}, \bibinfo{person}{Amit Sharma}, \bibinfo{person}{Hanna Wallach}, {and} \bibinfo{person}{Emine Yilmaz}.} \bibinfo{year}{2017}\natexlab{}.
\newblock \showarticletitle{Auditing Search Engines for Differential Satisfaction Across Demographics}. In \bibinfo{booktitle}{\emph{Proceedings of the 26th International Conference on World Wide Web Companion}} \emph{(\bibinfo{series}{WWW '17 Companion})}. \bibinfo{publisher}{International World Wide Web Conferences Steering Committee}, \bibinfo{pages}{626–633}.
\newblock


\bibitem[Moffat and Zobel(2008)]%
        {moffat2008rbp}
\bibfield{author}{\bibinfo{person}{Alistair Moffat} {and} \bibinfo{person}{Justin Zobel}.} \bibinfo{year}{2008}\natexlab{}.
\newblock \showarticletitle{Rank-biased precision for measurement of retrieval effectiveness}.
\newblock \bibinfo{journal}{\emph{ACM Transactions on Information Systems (TOIS)}} \bibinfo{volume}{27}, \bibinfo{number}{1} (\bibinfo{year}{2008}), \bibinfo{pages}{1--27}.
\newblock


\bibitem[Neelakanteswara et~al\mbox{.}(2024)]%
        {neelakanteswara2024ragstostyle}
\bibfield{author}{\bibinfo{person}{Abhiman Neelakanteswara}, \bibinfo{person}{Shreyas Chaudhari}, {and} \bibinfo{person}{Hamed Zamani}.} \bibinfo{year}{2024}\natexlab{}.
\newblock \showarticletitle{RAGs to Style: Personalizing LLMs with Style Embeddings}. In \bibinfo{booktitle}{\emph{Proceedings of the 1st Workshop on Personalization of Generative AI Systems (PERSONALIZE 2024)}}. \bibinfo{pages}{119--123}.
\newblock


\bibitem[Oosterhuis(2021)]%
        {oosterhuis2021PLRank}
\bibfield{author}{\bibinfo{person}{Harrie Oosterhuis}.} \bibinfo{year}{2021}\natexlab{}.
\newblock \showarticletitle{Computationally efficient optimization of plackett-luce ranking models for relevance and fairness}. In \bibinfo{booktitle}{\emph{Proceedings of the 44th International ACM SIGIR Conference on Research and Development in Information Retrieval}}. \bibinfo{pages}{1023--1032}.
\newblock


\bibitem[Oosterhuis(2022)]%
        {oosterhuis2022PLRank3}
\bibfield{author}{\bibinfo{person}{Harrie Oosterhuis}.} \bibinfo{year}{2022}\natexlab{}.
\newblock \showarticletitle{Learning-to-rank at the speed of sampling: Plackett-luce gradient estimation with minimal computational complexity}. In \bibinfo{booktitle}{\emph{Proceedings of the 45th International ACM SIGIR Conference on Research and Development in Information Retrieval}}. \bibinfo{pages}{2266--2271}.
\newblock


\bibitem[Plackett(1975)]%
        {plackett1975analysis}
\bibfield{author}{\bibinfo{person}{Robin~L Plackett}.} \bibinfo{year}{1975}\natexlab{}.
\newblock \showarticletitle{The analysis of permutations}.
\newblock \bibinfo{journal}{\emph{Journal of the Royal Statistical Society Series C: Applied Statistics}} \bibinfo{volume}{24}, \bibinfo{number}{2} (\bibinfo{year}{1975}), \bibinfo{pages}{193--202}.
\newblock


\bibitem[Raj and Ekstrand(2020)]%
        {raj2020comparing-fair-metrics}
\bibfield{author}{\bibinfo{person}{Amifa Raj} {and} \bibinfo{person}{Michael~D Ekstrand}.} \bibinfo{year}{2020}\natexlab{}.
\newblock \showarticletitle{Comparing fair ranking metrics}.
\newblock \bibinfo{journal}{\emph{arXiv preprint arXiv:2009.01311}} (\bibinfo{year}{2020}).
\newblock


\bibitem[Rashkin et~al\mbox{.}(2023)]%
        {rashkin-2023-ais}
\bibfield{author}{\bibinfo{person}{Hannah Rashkin}, \bibinfo{person}{Vitaly Nikolaev}, \bibinfo{person}{Matthew Lamm}, \bibinfo{person}{Lora Aroyo}, \bibinfo{person}{Michael Collins}, \bibinfo{person}{Dipanjan Das}, \bibinfo{person}{Slav Petrov}, \bibinfo{person}{Gaurav~Singh Tomar}, \bibinfo{person}{Iulia Turc}, {and} \bibinfo{person}{David Reitter}.} \bibinfo{year}{2023}\natexlab{}.
\newblock \showarticletitle{Measuring Attribution in Natural Language Generation Models}.
\newblock \bibinfo{journal}{\emph{Computational Linguistics}} \bibinfo{volume}{49}, \bibinfo{number}{4} (\bibinfo{date}{Dec.} \bibinfo{year}{2023}), \bibinfo{pages}{777--840}.
\newblock
\urldef\tempurl%
\url{https://doi.org/10.1162/coli_a_00486}
\showDOI{\tempurl}


\bibitem[Robertson et~al\mbox{.}(1995)]%
        {Robertson1995OkapiBM25}
\bibfield{author}{\bibinfo{person}{Stephen Robertson}, \bibinfo{person}{S. Walker}, \bibinfo{person}{S. Jones}, \bibinfo{person}{M.~M. Hancock-Beaulieu}, {and} \bibinfo{person}{M. Gatford}.} \bibinfo{year}{1995}\natexlab{}.
\newblock \showarticletitle{Okapi at TREC-3}. In \bibinfo{booktitle}{\emph{Proceedings of the Third Text REtrieval Conference}} \emph{(\bibinfo{series}{TREC-3})}. \bibinfo{publisher}{Gaithersburg, MD: NIST}, \bibinfo{pages}{109--126}.
\newblock


\bibitem[Ru et~al\mbox{.}(2024)]%
        {ru2024ragchecker}
\bibfield{author}{\bibinfo{person}{Dongyu Ru}, \bibinfo{person}{Lin Qiu}, \bibinfo{person}{Xiangkun Hu}, \bibinfo{person}{Tianhang Zhang}, \bibinfo{person}{Peng Shi}, \bibinfo{person}{Shuaichen Chang}, \bibinfo{person}{Cheng Jiayang}, \bibinfo{person}{Cunxiang Wang}, \bibinfo{person}{Shichao Sun}, \bibinfo{person}{Huanyu Li}, \bibinfo{person}{Zizhao Zhang}, \bibinfo{person}{Binjie Wang}, \bibinfo{person}{Jiarong Jiang}, \bibinfo{person}{Tong He}, \bibinfo{person}{Zhiguo Wang}, \bibinfo{person}{Pengfei Liu}, \bibinfo{person}{Yue Zhang}, {and} \bibinfo{person}{Zheng Zhang}.} \bibinfo{year}{2024}\natexlab{}.
\newblock \showarticletitle{{RAGC}hecker: A Fine-grained Framework for Diagnosing Retrieval-Augmented Generation}. In \bibinfo{booktitle}{\emph{The Thirty-eight Conference on Neural Information Processing Systems Datasets and Benchmarks Track}}.
\newblock
\urldef\tempurl%
\url{https://openreview.net/forum?id=J9oefdGUuM}
\showURL{%
\tempurl}


\bibitem[Saad-Falcon et~al\mbox{.}(2024)]%
        {saadfalcon2023ares}
\bibfield{author}{\bibinfo{person}{Jon Saad-Falcon}, \bibinfo{person}{Omar Khattab}, \bibinfo{person}{Christopher Potts}, {and} \bibinfo{person}{Matei Zaharia}.} \bibinfo{year}{2024}\natexlab{}.
\newblock \showarticletitle{{ARES}: An Automated Evaluation Framework for Retrieval-Augmented Generation Systems}. In \bibinfo{booktitle}{\emph{Proceedings of the 2024 Conference of the North American Chapter of the Association for Computational Linguistics: Human Language Technologies (Volume 1: Long Papers)}}, \bibfield{editor}{\bibinfo{person}{Kevin Duh}, \bibinfo{person}{Helena Gomez}, {and} \bibinfo{person}{Steven Bethard}} (Eds.). \bibinfo{publisher}{Association for Computational Linguistics}, \bibinfo{pages}{338--354}.
\newblock


\bibitem[Salemi et~al\mbox{.}(2024a)]%
        {salemi2024optimization}
\bibfield{author}{\bibinfo{person}{Alireza Salemi}, \bibinfo{person}{Surya Kallumadi}, {and} \bibinfo{person}{Hamed Zamani}.} \bibinfo{year}{2024}\natexlab{a}.
\newblock \showarticletitle{Optimization Methods for Personalizing Large Language Models through Retrieval Augmentation}. In \bibinfo{booktitle}{\emph{Proceedings of the 47th International ACM SIGIR Conference on Research and Development in Information Retrieval}} \emph{(\bibinfo{series}{SIGIR '24})}. \bibinfo{publisher}{Association for Computing Machinery}, \bibinfo{pages}{752–762}.
\newblock
\urldef\tempurl%
\url{https://doi.org/10.1145/3626772.3657783}
\showDOI{\tempurl}


\bibitem[Salemi et~al\mbox{.}(2024b)]%
        {salemi:lamp}
\bibfield{author}{\bibinfo{person}{Alireza Salemi}, \bibinfo{person}{Sheshera Mysore}, \bibinfo{person}{Michael Bendersky}, {and} \bibinfo{person}{Hamed Zamani}.} \bibinfo{year}{2024}\natexlab{b}.
\newblock \showarticletitle{{L}a{MP}: When Large Language Models Meet Personalization}. In \bibinfo{booktitle}{\emph{Proceedings of the 62nd Annual Meeting of the Association for Computational Linguistics (Volume 1: Long Papers)}}. \bibinfo{publisher}{Association for Computational Linguistics}, \bibinfo{pages}{7370--7392}.
\newblock


\bibitem[Salemi and Zamani(2024)]%
        {salemi2024erag}
\bibfield{author}{\bibinfo{person}{Alireza Salemi} {and} \bibinfo{person}{Hamed Zamani}.} \bibinfo{year}{2024}\natexlab{}.
\newblock \showarticletitle{Evaluating retrieval quality in retrieval-augmented generation}. In \bibinfo{booktitle}{\emph{Proceedings of the 47th International ACM SIGIR Conference on Research and Development in Information Retrieval}}. \bibinfo{pages}{2395--2400}.
\newblock


\bibitem[Sapiezynski et~al\mbox{.}(2019)]%
        {Sapiezynski19quantifying}
\bibfield{author}{\bibinfo{person}{Piotr Sapiezynski}, \bibinfo{person}{Wesley Zeng}, \bibinfo{person}{Ronald E~Robertson}, \bibinfo{person}{Alan Mislove}, {and} \bibinfo{person}{Christo Wilson}.} \bibinfo{year}{2019}\natexlab{}.
\newblock \showarticletitle{Quantifying the Impact of User Attentionon Fair Group Representation in Ranked Lists}. In \bibinfo{booktitle}{\emph{Companion Proceedings of The 2019 World Wide Web Conference}} \emph{(\bibinfo{series}{WWW '19})}. \bibinfo{publisher}{Association for Computing Machinery}, \bibinfo{pages}{553–562}.
\newblock


\bibitem[Saracevic(2016)]%
        {saracevic2016notion}
\bibfield{author}{\bibinfo{person}{Tefko Saracevic}.} \bibinfo{year}{2016}\natexlab{}.
\newblock \bibinfo{booktitle}{\emph{The Notion of Relevance in Information Science: Everybody knows what relevance is. But, what is it really?}}
\newblock \bibinfo{publisher}{Morgan \& Claypool Publishers}.
\newblock


\bibitem[Shrestha et~al\mbox{.}(2024)]%
        {Shrestha_2024_CVPR}
\bibfield{author}{\bibinfo{person}{Robik Shrestha}, \bibinfo{person}{Yang Zou}, \bibinfo{person}{Qiuyu Chen}, \bibinfo{person}{Zhiheng Li}, \bibinfo{person}{Yusheng Xie}, {and} \bibinfo{person}{Siqi Deng}.} \bibinfo{year}{2024}\natexlab{}.
\newblock \showarticletitle{FairRAG: Fair Human Generation via Fair Retrieval Augmentation}. In \bibinfo{booktitle}{\emph{Proceedings of the IEEE/CVF Conference on Computer Vision and Pattern Recognition (CVPR)}}. \bibinfo{pages}{11996--12005}.
\newblock


\bibitem[Singh and Joachims(2018)]%
        {singh2018fairness}
\bibfield{author}{\bibinfo{person}{Ashudeep Singh} {and} \bibinfo{person}{Thorsten Joachims}.} \bibinfo{year}{2018}\natexlab{}.
\newblock \showarticletitle{Fairness of exposure in rankings}. In \bibinfo{booktitle}{\emph{Proceedings of the 24th ACM SIGKDD international conference on knowledge discovery \& data mining}}. \bibinfo{pages}{2219--2228}.
\newblock


\bibitem[Singh and Joachims(2019)]%
        {singh_policylearning_fairness_2019}
\bibfield{author}{\bibinfo{person}{Ashudeep Singh} {and} \bibinfo{person}{Thorsten Joachims}.} \bibinfo{year}{2019}\natexlab{}.
\newblock \showarticletitle{Policy learning for fairness in ranking}.
\newblock In \bibinfo{booktitle}{\emph{Proceedings of the 33rd {International} {Conference} on {Neural} {Information} {Processing} {Systems}}}. Number 487. \bibinfo{publisher}{Curran Associates Inc.}, \bibinfo{pages}{5426--5436}.
\newblock


\bibitem[Tay et~al\mbox{.}(2023)]%
        {tay2023ul}
\bibfield{author}{\bibinfo{person}{Yi Tay}, \bibinfo{person}{Mostafa Dehghani}, \bibinfo{person}{Vinh~Q. Tran}, \bibinfo{person}{Xavier Garcia}, \bibinfo{person}{Jason Wei}, \bibinfo{person}{Xuezhi Wang}, \bibinfo{person}{Hyung~Won Chung}, \bibinfo{person}{Dara Bahri}, \bibinfo{person}{Tal Schuster}, \bibinfo{person}{Steven Zheng}, \bibinfo{person}{Denny Zhou}, \bibinfo{person}{Neil Houlsby}, {and} \bibinfo{person}{Donald Metzler}.} \bibinfo{year}{2023}\natexlab{}.
\newblock \showarticletitle{{UL}2: Unifying Language Learning Paradigms}. In \bibinfo{booktitle}{\emph{The Eleventh International Conference on Learning Representations}}.
\newblock
\urldef\tempurl%
\url{https://openreview.net/forum?id=6ruVLB727MC}
\showURL{%
\tempurl}


\bibitem[Yang and Stoyanovich(2017)]%
        {yang2017measuring}
\bibfield{author}{\bibinfo{person}{Ke Yang} {and} \bibinfo{person}{Julia Stoyanovich}.} \bibinfo{year}{2017}\natexlab{}.
\newblock \showarticletitle{Measuring fairness in ranked outputs}. In \bibinfo{booktitle}{\emph{Proceedings of the 29th international conference on scientific and statistical database management}}. \bibinfo{pages}{1--6}.
\newblock


\bibitem[Zamani and Bendersky(2024)]%
        {zamani24stochasticRAG}
\bibfield{author}{\bibinfo{person}{Hamed Zamani} {and} \bibinfo{person}{Michael Bendersky}.} \bibinfo{year}{2024}\natexlab{}.
\newblock \showarticletitle{Stochastic RAG: End-to-End Retrieval-Augmented Generation through Expected Utility Maximization}. In \bibinfo{booktitle}{\emph{Proceedings of the 47th International ACM SIGIR Conference on Research and Development in Information Retrieval}} \emph{(\bibinfo{series}{SIGIR '24})}. \bibinfo{publisher}{Association for Computing Machinery}, \bibinfo{pages}{2641–2646}.
\newblock
\urldef\tempurl%
\url{https://doi.org/10.1145/3626772.3657923}
\showDOI{\tempurl}


\bibitem[Zamani et~al\mbox{.}(2022)]%
        {zamani:reml}
\bibfield{author}{\bibinfo{person}{Hamed Zamani}, \bibinfo{person}{Fernando Diaz}, \bibinfo{person}{Mostafa Dehghani}, \bibinfo{person}{Donald Metzler}, {and} \bibinfo{person}{Michael Bendersky}.} \bibinfo{year}{2022}\natexlab{}.
\newblock \showarticletitle{Retrieval-Enhanced Machine Learning}. In \bibinfo{booktitle}{\emph{Proceedings of the 45th Annual International ACM SIGIR Conference on Research and Development in Information Retrieval}}.
\newblock


\bibitem[Zehlike et~al\mbox{.}(2017)]%
        {zehlike2017fair}
\bibfield{author}{\bibinfo{person}{Meike Zehlike}, \bibinfo{person}{Francesco Bonchi}, \bibinfo{person}{Carlos Castillo}, \bibinfo{person}{Sara Hajian}, \bibinfo{person}{Mohamed Megahed}, {and} \bibinfo{person}{Ricardo Baeza-Yates}.} \bibinfo{year}{2017}\natexlab{}.
\newblock \showarticletitle{Fa* ir: A fair top-k ranking algorithm}. In \bibinfo{booktitle}{\emph{Proceedings of the 2017 ACM on Conference on Information and Knowledge Management}}. \bibinfo{pages}{1569--1578}.
\newblock


\bibitem[Zhang et~al\mbox{.}(2024b)]%
        {Zhang24llmutilityjudgment}
\bibfield{author}{\bibinfo{person}{Hengran Zhang}, \bibinfo{person}{Ruqing Zhang}, \bibinfo{person}{Jiafeng Guo}, \bibinfo{person}{Maarten de Rijke}, \bibinfo{person}{Yixing Fan}, {and} \bibinfo{person}{Xueqi Cheng}.} \bibinfo{year}{2024}\natexlab{b}.
\newblock \showarticletitle{Are Large Language Models Good at Utility Judgments?}. In \bibinfo{booktitle}{\emph{Proceedings of the 47th International ACM SIGIR Conference on Research and Development in Information Retrieval}} \emph{(\bibinfo{series}{SIGIR '24})}. \bibinfo{publisher}{Association for Computing Machinery}, \bibinfo{pages}{1941–1951}.
\newblock


\bibitem[Zhang et~al\mbox{.}(2024a)]%
        {zhang2024diffusedistributionslanguage}
\bibfield{author}{\bibinfo{person}{Yiming Zhang}, \bibinfo{person}{Avi Schwarzschild}, \bibinfo{person}{Nicholas Carlini}, \bibinfo{person}{J~Zico Kolter}, {and} \bibinfo{person}{Daphne Ippolito}.} \bibinfo{year}{2024}\natexlab{a}.
\newblock \showarticletitle{Forcing Diffuse Distributions out of Language Models}. In \bibinfo{booktitle}{\emph{First Conference on Language Modeling}}.
\newblock
\urldef\tempurl%
\url{https://openreview.net/forum?id=9JY1QLVFPZ}
\showURL{%
\tempurl}


\end{thebibliography}
